\documentclass[reprint,amsmath,amssymb,apsm,siunitx,twocolumn]{revtex4-2}
\usepackage[utf8]{inputenc}
\usepackage{bbm}
\usepackage{natbib}
\usepackage{color}
\usepackage{wasysym}
\usepackage{subfloat}
\usepackage{graphicx}
\usepackage{adjustbox} 
\usepackage{dcolumn}
\usepackage{breqn}
\usepackage{physics}
\usepackage{float}
\usepackage{soul}
\usepackage{algorithmic}
\usepackage[toc,page]{appendix}
\usepackage{tabularx}
\usepackage{todonotes}
\usepackage{verbatim}
\usepackage{ragged2e}
\usepackage{siunitx}
\usepackage{steinmetz}
\usepackage{makecell, booktabs, multirow}

\usepackage{lipsum, fancyvrb}
\usepackage[caption=false]{subfig} 
\usepackage{xcolor}
\makeatletter
\renewcommand*{\fnum@figure}{{\normalfont\bfseries \figurename~\thefigure}}
\renewcommand*{\@caption@fignum@sep}{\textbf{ : }}
\makeatother
\makeatletter
\renewcommand*{\fnum@table}{{\normalfont\bfseries \tablename~\thetable}}
\makeatletter

\usepackage{mathtools}

\DeclarePairedDelimiterX\set[1]\lbrace\rbrace{\def\given{\;\delimsize\vert\;}#1}

\begin{document}

\title{Combined tools for Particle-In-Cell simulations performed with transversely asymmetric chirped lasers}

\author{I. Moulanier, F. Massimo, T. L. Steyn, B. Cros}
\affiliation{LPGP$,$ CNRS$,$ Universit$\text{é}$ Paris Saclay$,$ 91405 Orsay$,$ France$,$ \break \rm{Corresponding author : I. Moulanier}$,$ \rm{ioaquin.moulanier@universite-paris-saclay.fr}}

\begin{abstract}
    We introduce the Asymmetric Chirped Electric field reconstruction (ACE) toolbox, a suite of algorithms enabling the reconstruction of a laser transverse distribution coupled to a chirped temporal profile. This suite includes the implementation of the reconstructed distribution in Particle In-Cell simulations in cylindrical and Cartesian geometry. Under the assumption of negligible spatio-temporal couplings, the ACE toolbox extends a previously established method for realistic simulations of the transverse distribution by adding spectral chirping to the modal decomposition of the electric field. 
    The relevance of the ACE toolbox is demonstrated through the modelling of a Laser Wakefield Acceleration experiment where the produced electron bunch was optimized via spectral chirping of the laser pulse. 
\end{abstract}

\maketitle

\section{Introduction}
Laser WakeField Acceleration (LWFA) is a particle acceleration technique in which an ultra-high intensity laser propagates in an under-dense plasma. Plasma waves generated in the wake of the laser can sustain accelerating gradients exceeding the current technological limits of accelerators based on RF cavities~\cite{TajimaDawson1979, Esarey2009}. 
However, improvements in controlling laser parameters are necessary to achieve stable electron beams~\cite{assmann2020eupraxia} with low energy dispersion~\cite{andre2018control}.

State-of-the-art experiments and simulations show correlations between the shot-to-shot transverse distribution of the laser driver and properties of the accelerated electrons~\cite{popp2010all, ferri2016effect, maier2020decoding, Dickson2022,Dickson2023,moulanier2023modeling}.
Due to this dependence on the laser field distribution, simulations of LWFA experiments require accurate modelling of the laser electric field in order to reproduce the experimental electron bunch properties in Particle in Cell (PIC)~\cite{BirdsallLangdon} simulations.
The transverse field distribution can be reconstructed from fluence measurements at transverse planes on the laser propagation axis. The use of laser fields reconstructed from fluence measurements for detailed PIC modelling was demonstrated in~\cite{ferri2016effect} with 3D PIC simulations and in~\cite{zemzemi2020high} with PIC simulations in cylindrical geometry with azimuthal mode decomposition~\cite{lifschitz2009particle}. In these references, field reconstruction was performed through the Gerchberg-Saxton Algorithm (GSA) \cite{gerchberg1972practical}. More recently~\cite{Dickson2022, moulanier2023modeling}, the reconstructed fields in PIC simulations were obtained through the Gerchberg-Saxton Algorithm with Mode Decomposition (GSA-MD), described in~\cite{moulanier2023fast, massimo2025laser}. Moreover, in~\cite{Dickson2022, moulanier2023modeling, thevenet2025lasy} it has been shown that the laser field reconstructed with the GSA-MD can be easily integrated in PIC simulations with cylindrical geometry and azimuthal mode decomposition, enabling an excellent agreement between experimental results and PIC simulations, with computational resources considerably smaller than those needed by full 3D simulations. In~\cite{Dickson2022,moulanier2023modeling}, this agreement with experimental results is considerably higher than the one of PIC simulations with idealized Gaussian beams used as laser model.

\textcolor{black}{Controlled spatio-temporal perturbations of high power laser systems~\cite{popp2010all, sainte2017controlling, leroux2020description} have been proposed to improve the characteristics of electrons accelerated by LWFA.} 
In particular, several experiments have demonstrated that a fine tuning of the laser temporal profile by spectral chirping affects the distribution of the injected electron bunch~\cite{leemans2002electron, schroeder2003frequency, toth2003tuning, kalmykov2012laser, kim2017stable, pathak2018spectral, grigoriadis2023efficient}.
\textcolor{black}{Due to the large number of input laser and plasma parameters for the optimization of the electron bunch characteristics, Bayesian Optimization (BO)~\cite{Jones1998,Frazier2018} is frequently used for fine tuning of PIC simulations and experiments of LWFA~\cite{kirchen2021optimal, jalas2021bayesian, ferran2023bayesian, jalas2023tuning, irshad2023multi, irshad2024pareto, valenta2025bayesian}. 
The addition of chirp coefficients to BO of laser driven sources~\cite{shalloo2020automation, loughran2023automated, glenn2026characterization, zeraouli2026exploring}} has demonstrated that optimal working points can be found by modulating the laser spectral phase up to the fourth order in frequency, thus optimizing the particle bunch parameters.
In an experiment described in~\cite{Dickson2023, Backhouse2025}, BO was used at the Lund Laser Centre (LLC) to find  the optimal laser focal plane position and phase chirping coefficients that obtain low energy spread electron bunches. To reproduce the optimal working point described in~\cite{Backhouse2025} with PIC simulations, a reconstruction of the laser field was used with information from fluence measurements at transverse planes and measurements of the spectral phase.

In this article, we describe a toolbox for the reconstruction of an Asymmetric Chirped Electric field (ACE), suitable for PIC simulations of laser-plasma interactions. It combines accurate reconstruction of the electric field temporal chirp and transverse asymmetry in the form of a modal decomposition using paraxial solutions of the wave equation. The ACE toolbox can be used for the modelling of LWFA in regimes where laser spatio-temporal couplings (STC)~\cite{Akturk_2010, Pariente2016, Jeandet22} are negligible.

\begin{figure}[h]
	\centering
	  \includegraphics[width=1.0\linewidth]{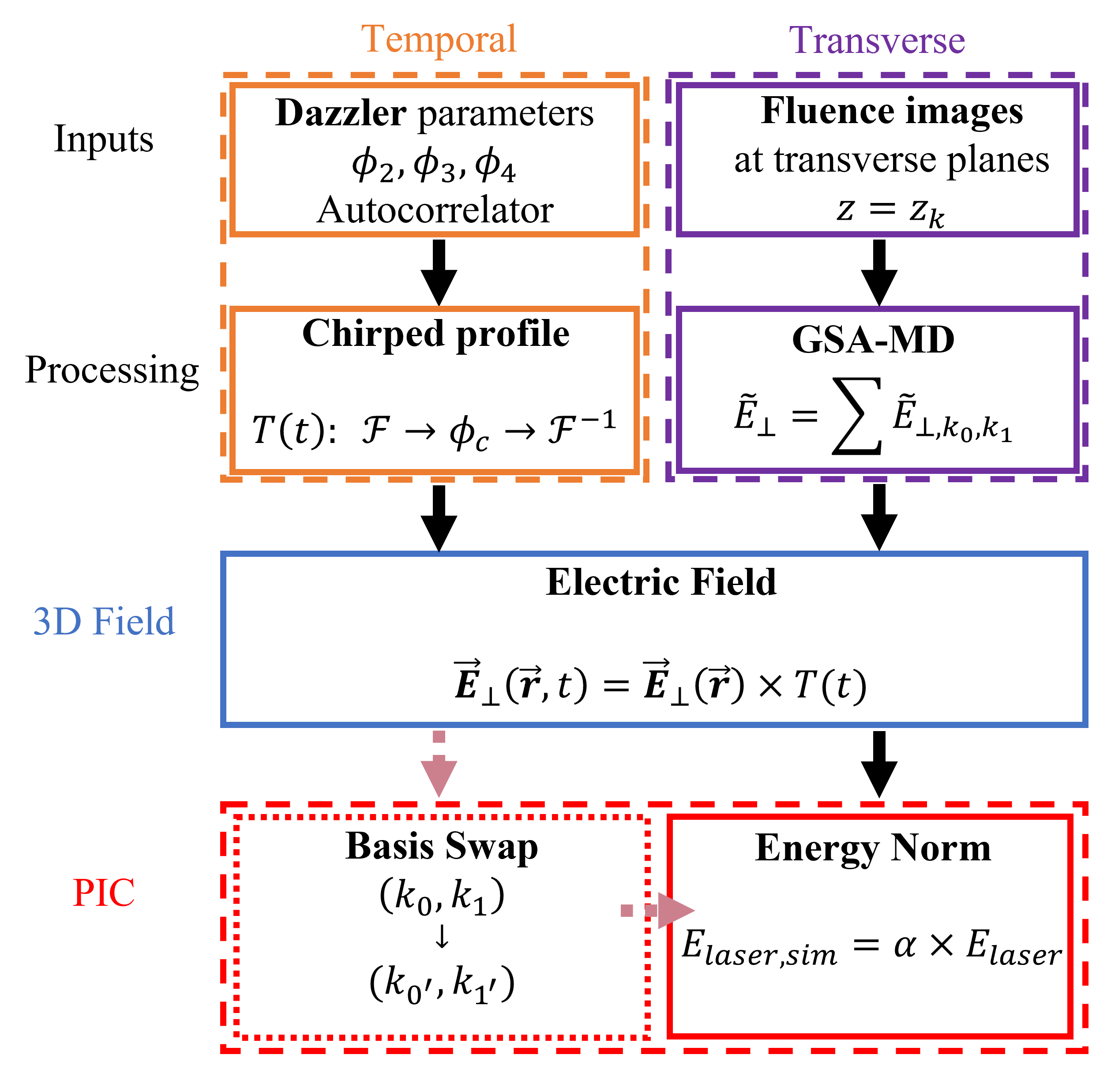}
	\caption{Schematic representation of the ACE toolbox for PIC simulations with a transversely asymmetric chirped laser. The ``Basis Swap" step written within a dotted frame indicates the optional operation that depends on the geometry of the simulation.
	This step was applied to the FBPIC simulation of the use case presented in this paper.
    Each frame included in this diagram is described in order of apparition in the sections of this article.}
	\label{fig:Framework}
\end{figure}

The method for the 3D laser reconstruction described in this paper is depicted in Fig.~\ref{fig:Framework}. The inputs are laser characteristics which can be measured experimentally: chirp coefficients $\phi_2,~\phi_3,~\phi_4$ that shape the spectral phase, and fluence distribution at transverse planes with coordinates $z_k$ on the laser propagation axis in vacuum. These data are processed to retrieve the temporal profile of the laser through an inverse Fourier transform and a paraxial modal decomposition of the transverse field distribution through the GSA-MD. These field reconstructions in the longitudinal and transverse directions are combined to obtain the 3D-reconstructed laser electric field. The pulse energy is scaled as a function of the measured energy and decomposed in the modal basis used in PIC simulations in cylindrical geometry with azimuthal mode decomposition.

As use case for the ACE toolbox, the operations shown in Fig.~\ref{fig:Framework} and described in this paper are performed using laser data collected at the LLC for the experiment described in~\cite{Dickson2023, Backhouse2025}. 
The considered \si{TW}-class laser system is characterized by a total energy $E_{laser}=0.87$ \si{J}, a $1/e^2$ temporal duration $\tau_l=33.3$ \si{\femto\second}, and a $1/e^2$ waist at focus $w_0=12.1$ \si{\micro\metre}.
The laser was focused at the entrance of a gas cell filled with a mixture of $95\,\si{\percent}$ $\mathrm{H_2}$ and $5\,\si{\percent}$ of $\mathrm{N_2}$ to realize ionization injection~\cite{pak2010injection, mcguffey2010ionization, chen2012theory, mirzaie2015demonstration} of electrons in the plasma wakefield. Automatic shot-to-shot BO was performed to significantly improve the output electron bunch quality in terms of charge, energy gain and energy spread. For this procedure, four main laser parameters were optimized by the BO: the polynomial chirp coefficients $\phi_2$, $\phi_3$, $\phi_4$, and the position of the focal plane of the laser within the gas cell compartment, $z_{f}$.

This paper is organized as follows: the second section presents the methodology for the 3D (transverse and temporal) reconstruction of the laser electric field based on fluence and spectral phase measurements. It shows how the fully reconstructed laser electric field distribution can be integrated in FBPIC~\cite{Lehe2016}, a PIC code with azimuthal mode decomposition \cite{lifschitz2009particle} and suitable for the modelling of LWFA. 
The third section presents PIC simulations in the conditions of the optimal working point found in the LLC experiment described in~\cite{Backhouse2025}.
Finally, the effects of second order spectral chirping on the characteristics of the electron bunches are studied through PIC simulations where the spectral phase parameters are changed around that working point. \textcolor{black}{Appendix\,\ref{sec:appendix_GSAMD} reports the parameters of GSA-MD used for the PIC simulations in this article. Appendix\,\ref{sec:appendix_transverse_distribution} describes the modelling of laser transverse distribution with the ACE toolbox. Appendix\,\ref{sec:appendix_sims} reports the numerical parameters used in PIC simulations of this article. Appendix\,\ref{sec:appendix_fit_E} shows the fitted scaling of the energy of the electron bunches as function of $\phi_2$ obtained in the PIC simulations. Appendix\,\ref{sec:appendix_fit_sigmaE} shows the scaling of the charge and energy spread obtained in the simulations.}

\section{Spatio-temporal reconstruction of a chirped laser}\label{sec:Methodology}

In vacuum, Gaussian-like laser pulses can be expressed as a superposition of individual solutions of the paraxial wave equation.
Neglecting spatio-temporal couplings (STC) in each individual mode and assuming they are temporally synchronized, i.e. centered at the same co-moving coordinate $\xi_0=z_0-t_0$, the total electric field near focus is also a solution of the paraxial equation with no STC~\cite{pierce2023arbitrarily}. 
The LLC laser characteristic transverse size at focus is $w_0=12.1$ \si{\micro\metre} and its unchirped longitudinal size is $c\tau_l=10.0$ \si{\micro\metre}. Both characteristic lengths are considerably larger than $\lambda_0=0.8$ \si{\micro\metre} and fulfill the slowly varying envelope condition of the paraxial approximation~\cite{svelto2010principles}.

\textcolor{black}{Without STC, }in Cartesian and polar coordinates, such laser field can be written as the real part of the product of a transverse spatial distribution with a temporal profile that contains the sinusoidal oscillations of the electric field. In the following, without loss of generality for the reconstruction method, we assume linear polarization. In the focal plane, the transverse component of the electric field, $E_{\perp}$, can be written as~\cite{quesnel1998theory, pierce2023arbitrarily}:

\begin{equation}\label{eq:Efield_3D}
    E_{\perp}(\overrightarrow{r}, z=z_f, t) = \real \{ \tilde{E}_{\perp}(\overrightarrow{r}, z=z_f)\times \tilde{T}(t)\},
\end{equation}
where $E_{\perp}(\overrightarrow{r}, z=z_f, t)$ is the linearly polarized electric field distribution at focus, $\real$ the Real Part operator, $\tilde{E}_{\perp}(\overrightarrow{r}, z)$ the spatial complex component of the electric field and $\tilde{T}(t)$ its complex temporal profile.
The spatio-temporal decoupling allows to separate the retrieval of the temporal and transverse contributions in the 3D electric field distribution. This is illustrated in Fig.\,\ref{fig:Framework} ``Temporal" and ``Transverse" dashed sections, where the ``Input" and ``Processing" steps that are necessary to the ``3D Reconstruction" [Eq.\,(\ref{eq:Efield_3D})] are performed in parallel for both distributions.

These two steps are described in subsections \ref{sec:Temporal_reconstruction} [$\tilde{T}(t)$ in Eq.\,(\ref{eq:Efield_3D})] and \ref{sec:Transverse_reconstruction} [$\tilde{E}_{\perp}$ in Eq.\,(\ref{eq:Efield_3D})], while the modelling of $E_{\perp}(\overrightarrow{r}, z, t)$ in a PIC code will be discussed in section \ref{sec:Implementation}.

\subsection{Temporal reconstruction}\label{sec:Temporal_reconstruction}

In the following we assume that the temporal profile without chirp is a Gaussian profile centered with a carrier frequency $\omega_0$. Under controlled modulations of its temporal phase, or chirp, the temporal profile $\tilde{T}(t)$ is written explicitly as:
\begin{equation}\label{eq:Efield_time}
    \tilde{T}(t) = \mathcal{F}^{-1}\biggl\{\mathcal{F}\biggl\{\mathrm{\exp}\left[-\dfrac{t^2}{\tau^2}\right]\exp\left[i\omega_0 t\right]\biggr\}\mathrm{\exp}\left[i\phi_{c}(\omega)\right]\biggr\},
\end{equation}
where $\mathcal{F}$, $\mathcal{F}^{-1}$ are the Fourier transform and inverse Fourier transform operators, $\omega_0=2\pi c/\lambda_0$ is the carrier frequency of the laser, $\tau$ its $1/e^2$ duration and $\phi_{c}(\omega$) the additional spectral phase. In experiment, chirp coefficients can be controlled to a high degree by propagating the laser in a nonlinear medium excited by an acousto-optic modulator~\cite{tournois1997acousto, verluise2000amplitude}.
In the Fourier space, the parameter $\phi_{c}(\omega)$ is generated by the modulations of the chirp coefficients. Close to the value of the central frequency $\omega_0$, the additional chirped phase is represented by a Taylor expansion as a function of the frequency $\omega$: 

\begin{equation}\label{eq:phi_chirp}
    \phi_{c}(\omega) = \sum_{n=0}^{+\infty} \phi_n\frac{(\omega - \omega_0)^n}{n!},
\end{equation}
where $\phi_n$ is the chirp coefficient of order $n$. Each expansion coefficient $\phi_n$ acts differently on both the shape of the temporal laser envelope, and the distribution of instantaneous frequencies defining the laser oscillations. In this article, the optimization and effects of $\phi_0$ and $\phi_1$ are not investigated as the former changes the laser carrier envelope phase and the latter adds delay to the laser temporal profile. The effects of $\phi_0$ in particular are notable in cases of few-cycle lasers~\cite{huijts2021identifying}, which is not the case here ($\lambda_0/c=0.08\tau_l$).
The effects of chirp coefficients up to $n=2$ are predicted by analytically solving Eq.\,(\ref{eq:Efield_time}) for the spectral phase $\phi_c(\omega)$ given by Eq.\,(\ref{eq:phi_chirp}).
For a Gaussian temporal profile with a chirp coefficient $\phi_2=\partial^2 \phi_{c}(\omega)/\partial\omega^2|_{\omega=\omega_0} \neq 0$, the chirped duration $\tau_{l,c}$ is larger than the unchirped one, $\tau_l$~\cite{borzsonyi2013we, mohseni2018resolution}:
\begin{equation}\label{eq:tau_chirp}
    \tau_{l,c}=\tau_{l}\sqrt{1+\dfrac{4\phi_2^2}{\tau_{l}^4}}.
\end{equation}
The normalized electric field chirped amplitude, $a_{0,c}$, is reduced compared to the unchirped one, $a_0$:
\begin{equation}\label{eq:a_chirp}
        a_{0, c} = a_{0}\left(1+\dfrac{4\phi_2^2}{\tau_{l}^4}\right)^{-1/4},
\end{equation}
where $a_0=\mathrm{e}E_0/(m_e c\omega_0)$, with $\mathrm{e}$ the elementary charge, $m_e$ the electron mass and $E_0$ the maximum amplitude of the laser electric field in vacuum.
For $n \geq 3$, solutions of Eq.\,(\ref{eq:Efield_time}) can be numerically computed using the numerical Fast Fourier Transform (FFT) and the Inverse Fast Fourier Transform (IFFT)~\cite{cooley1965algorithm}. The effects of the third order coefficient $\phi_3=\partial^3 \phi_{c}(\omega)/\partial\omega^3|_{\omega=\omega_0}$ on the temporal envelope and frequencies are well documented for ultra-short lasers~\cite{diels2006ultrashort, mohseni2018resolution, jolly2019spectral}.

\begin{figure}[h]
	\centering
	  \includegraphics[width=1.0\linewidth]{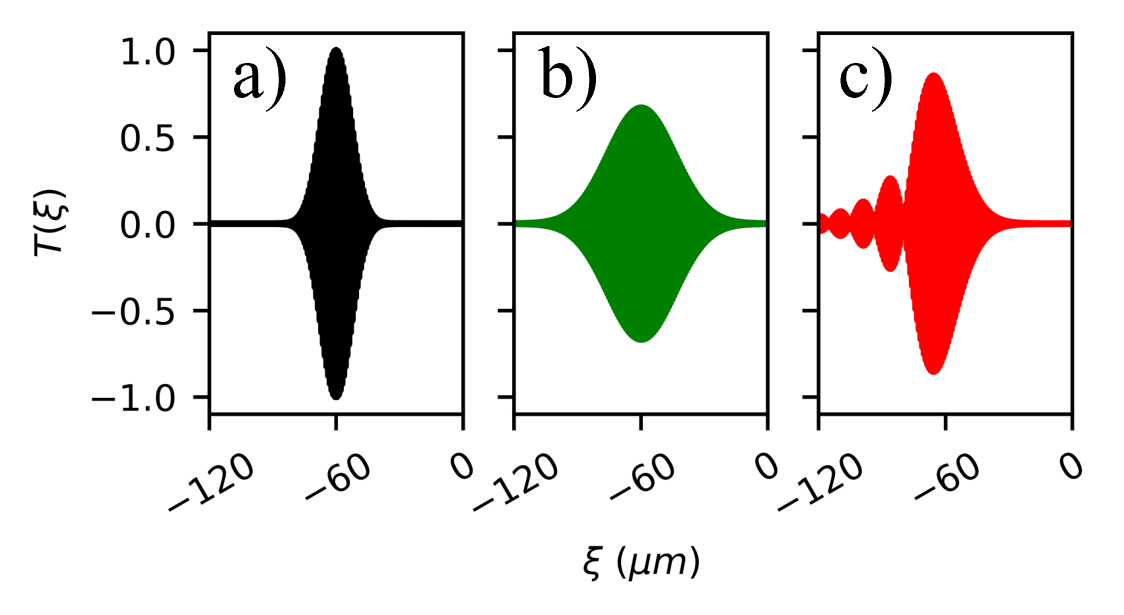}
	\caption{Effects of chirp coefficients on the laser electric field temporal profile, $T(\xi)=\real\{\tilde{T}(\xi)\}$, where $\xi=z-ct$, as predicted by Eqs.\,(\ref{eq:Efield_time}) and (\ref{eq:phi_chirp}):
	a) Unchirped Gaussian profile with a duration $\tau_l=33.3$ \si{\femto\second} and $\phi_c(\omega)=0$ - b) Gaussian profile with $\tau_l=33.3$ \si{\femto\second} for $\phi_2=\tau_l^2$ ($\phi_{n\neq2}=0$), resulting in $\tau_{l,c}=74.5$ \si{\femto\second} - c) Gaussian profile with  $\tau_l=33.3$ \si{\femto\second} and $\phi_3=\tau_l^3$ ($\phi_{n\neq3}=0$).
	}
	\label{fig:Chirp_illustration}
\end{figure}

The temporal profile of lasers chirped to the second and third order are illustrated in Fig.\,\ref{fig:Chirp_illustration}. For significant values of $\phi_2$, the laser duration is enlarged and its peak amplitude is diminished [Fig.\,\ref{fig:Chirp_illustration}.(b)]. 
Since chirp coefficients preserve the total energy, the decrease in peak amplitude is directly correlated to the increase of laser duration. As seen in Eqs.\,(\ref{eq:tau_chirp}) and (\ref{eq:a_chirp}), both are impacted by a chirp factor $\chi(\phi_2)=\left(1+4\phi_2^2/\tau_{l}^4\right)\geq1$, which ensures the conservation of the energy scaling $a_0^2\tau_l=a^2_{0,c}\tau_{l,c}$.
For significant positive (resp. negative) values of $\phi_3$ and $\phi_2=0$, the laser temporal profile is asymmetric with a longer (shorter) front half, and exhibits multiple sub-bunches at the back (front) half [Fig.\,\ref{fig:Chirp_illustration}.(c)].

In the experiment described in~\cite{Backhouse2025} the additional chirp coefficients were applied to the laser temporal profile using a DAZZLER~\cite{kaplan2002dazzler}, which recorded the set values of $\phi_n$. Based on Eq.\,(\ref{eq:phi_chirp}), substituting the chirp coefficients $\phi_n$ by the non-zero values set by the DAZZLER allows a full description of the additional chirped phase.

For the temporal profile reconstruction, the spectral phase is added to the initially unchirped Gaussian temporal profile in the Fourier domain as written in Eq.\,(\ref{eq:Efield_time}). During this step, the operator $\mathcal{F}$ is substituted by the numerical FFT in Eq.\,(\ref{eq:Efield_time}). The FFT requires a sufficiently high sampling rate in comparison to the signal central wavelength $\lambda_0=0.8$ \si{\micro\metre}. To calculate the numerical FFT, the discretized time interval over which the integral is performed has a sampling period $\mathop{dt}=\lambda_0/1000/c$. The sampling rate $1/\mathop{dt}$ fulfills the Nyquist sampling criterion since $1/\mathop{dt}\gg2c/\lambda_0$~\cite{shannon2006communication}. The chirped temporal profile is retrieved by substituting the inverse transform $\mathcal{F}^{-1}$ with the numerical IFFT in Eq.\,(\ref{eq:Efield_time}).

\subsection{Transverse reconstruction}\label{sec:Transverse_reconstruction}

Close to the focal plane in vacuum where the paraxial hypothesis is satisfied, the 2D spatial component of Eq.\,(\ref{eq:Efield_3D}), $\tilde{E}_{\perp}(\overrightarrow{r}, z)$, can be \textcolor{black}{approximated} as a truncated sum of eigenmodes of the paraxial wave equation in vacuum. The most appropriate choice of the selected basis --- Hermite-Gauss (HG) in Cartesian coordinates~\cite{siegman1973hermite} or Laguerre-Gauss (LG) in cylindrical coordinates~\cite{allen1992orbital} --- directly depends on the geometry of the input fluence images and that of the code in which the laser is to be modelled.
In both cases, $\tilde{E}_{\perp}$ is expressed as a sum of eigenmodes of two transverse coordinates:
\begin{equation}\label{eq:Efield_spatial}
    \tilde{E}_{\perp}(\overrightarrow{r}, z) = \sum_{k_0,\,k_1=0}^{N_{0},\,N_{1}} \tilde{\mathrm{E}}_{\perp, k_{0}, k_{1}}(\overrightarrow{r}, z) 
\end{equation}
\begin{equation}\label{eq:Efield_spatial_2}
    \begin{aligned}
    &\iint \tilde{\mathrm{E}}_{\perp, k_{0}, k_{1}}\tilde{\mathrm{E}}^*_{\perp, k_{0'}, k_{1'}} \mathrm{det}[\mathbf{J}(r_0,r_1)] \mathop{dr_0} \mathop{dr_1} = \tilde{C}_{k_{0},k_{1}}\tilde{C}^*_{k_{0'},k_{1'}}\\ &\hspace{180pt} \times\delta_{k_{0},k_{0'}}\delta_{k_{1}, k_{1'}}, 
    \end{aligned}
\end{equation}
where $\overrightarrow{r}$ is the radial vector in the plane perpendicular to the laser propagation axis, $k_{0,1}$ the eigenmode indices for the transverse plane axes, $N_{0,1}$ their respective maximum eigenmodes, $r_{0,1}$ their transverse coordinates (components of $\overrightarrow{r}$), $\mathrm{det}[\mathbf{J}(r_0,r_1)]$ the determinant of the Jacobian matrix from the Cartesian coordinates ($x$, $y$) to ($r_0$, $r_1$), $\tilde{C}_{k_{0},k_{1}}$ the complex amplitude of the mode $\tilde{\mathrm{E}}_{\perp,k_0,k_1}$ and $\delta_{k_0,k_0'}$ the Kroenecker symbol which equals $1$ if $k_0=k_0'$ and $0$ otherwise. The equations using the notations $r_{0,1}$ are in common for the Cartesian geometry ($r_0=x$, $r_1=y$, $\mathrm{det}[\mathbf{J}]=1$) and cylindrical geometry ($r_0=r$, $r_1=\theta$, $\mathrm{det}[\mathbf{J}]=r$).

\textcolor{black}{A usual experimental characterization for $\tilde{E}_{\perp}$ is given by diagnostics of the fluence $F(\overrightarrow{r},z)$, which is the transverse distribution of the laser energy per unit surface. In the absence of STC, using the electric field definition of Eq.\,(\ref{eq:Efield_3D}), the fluence integral is given by :
\begin{equation}\label{eq:fluence}
    F(\overrightarrow{r},z)=\textcolor{black}{\frac{\varepsilon_0c}{2}}|\tilde{E}_{\perp}|^2(\overrightarrow{r}, z)\int |\tilde{T}|^2(t) \mathop{dt},
\end{equation}
where $\varepsilon_0$ is the vacuum permittivity. \textcolor{black}{Under this approximation, the energy integral of $F(\overrightarrow{r},z)$ is directly linked to the orthogonality of the eigenmodes in Eq.\,(\ref{eq:Efield_spatial_2}):
\begin{equation}\label{eq:energy}
    \iint F(\overrightarrow{r},z)\mathrm{det}[\mathbf{J}(r_0,r_1)] \mathop{dr_0} \mathop{dr_1} = \mathcal{T}\times\sum_{k_0,k_1=0}^{N_{0},N_{1}} |\tilde{C}_{k_0,k_1}|^2,
\end{equation}
}
where $\mathcal{T}=\varepsilon_0c/2\int |\tilde{T}|^2(t) \mathop{dt}$.
By definition, $F(\overrightarrow{r},z)$ is real-valued. Therefore, with knowledge of the laser temporal profile, individual images of the laser fluence can at best give information on $|\tilde{E}_{\perp}|$ at their corresponding position on the laser propagation axis.
To get the complex field $\tilde{E}_{\perp}$, including its complex phase, it is necessary to retrieve its wavefront represented by the phasemap $\psi(\overrightarrow{r})=\arctan[\mathfrak{Im}\{\tilde{E}_{\perp}\}/\real\{\tilde{E}_{\perp}\}]$, where $\mathfrak{Im}$ is the Imaginary Part operator.}

The phasemap can be reconstructed with GSA-like algorithms~\cite{gerchberg1972practical}. These procedures require the knowledge of the modulus of the electric field in at least two planes. An iterative loop retrieves the complex phase that allows the light to travel from the first plane to the others using the paraxial approximation.

\textcolor{black}{The original formulation of the GSA requires a set of two planes (source and image), e.g. $\{z_0$, $z_1\}$, in which the transverse distribution $|\tilde{E}_{\perp}|$ is known. At the first iteration of a loop over the measurement planes, the reconstructed electric field is initialized as:
\begin{equation}\label{eq:GSA_step1}
    \tilde{E}_{\perp, \mathrm{GSA}}(\overrightarrow{r}, z_0) = |\tilde{E}_{\perp}(\overrightarrow{r}, z_0)|\exp\left[i \psi_{0}(\overrightarrow{r})\right],
\end{equation}
where $|\tilde{E}_{\perp}|$ is the measured electric field modulus and $\psi_{0}(\overrightarrow{r})$ the initial arbitrary transverse phase, e.g. $\psi_{0}(\overrightarrow{r})=\overrightarrow{0}$. This initial electric field is then propagated to $z_1$, using e.g. a Fourier Transform as in the original article introducing the GSA~\cite{gerchberg1972practical}, or a Fresnel transform $\mathcal{F_R}$~\cite{born2013principles}:
\begin{equation}\label{eq:GSA_step2}
    \tilde{E}_{\perp, \mathrm{GSA}}(\overrightarrow{r}, z_1) = \mathcal{F_R}\{ \tilde{E}_{\perp, \mathrm{GSA}}(\overrightarrow{r}, z_0) \}(\overrightarrow{r}, z_1)
\end{equation}
The convergence of the GSA is ensured by forcing the propagated electric field to match the modulus of the measured distribution in $z_1$, yielding the following new expression in $z_1$:
\begin{equation}\label{eq:GSA_step3}
    \tilde{E}_{\perp, \mathrm{GSA}, \mathrm{new}}(\overrightarrow{r}, z_1) = |\tilde{E}_{\perp}(\overrightarrow{r}, z_1)|\exp\left[i \psi_{\mathrm{GSA}}(\overrightarrow{r})\right],
\end{equation}
where $\psi_{\mathrm{GSA}}(\overrightarrow{r})=\arctan[\mathfrak{Im}\{\tilde{E}_{\perp, \mathrm{GSA}}\}/\real\{\tilde{E}_{\perp, \mathrm{GSA}}\}]$ is the phase of the GSA electric field calculated in Eq.\,(\ref{eq:GSA_step2}). Then, the field found with Eq.\,(\ref{eq:GSA_step3}) is back-propagated to $z_0$, where the measured modulus of the field at that plane is imposed, and the loop over the measurement planes that progressively improves the reconstruction of the phase distribution can restart.
GSA-like algorithms use variations of the main steps described earlier and in Eqs.\,(\ref{eq:GSA_step1}-\ref{eq:GSA_step3}).}

\textcolor{black}{Within the ACE toolbox, the reconstruction of the laser from different fluence images along the propagation axis is done using the Gerchberg-Saxton Algorithm with Mode Decomposition (GSA-MD)~\cite{moulanier2023fast, massimo2025laser}.
The main differences with the original GSA are that more than 2 planes are used for the reconstruction, and the electric field is described as the modal sum as in Eq.\,(\ref{eq:Efield_spatial}). The propagation from one plane $z_k$ to the next $z_{k+1}$ is done in the GSA-MD by using the analytical expressions of either LG or HG mode basis, as shown in~\cite{Sroor2021}.}

As defined in Eq.\,(\ref{eq:fluence}), without STC, the chirped temporal profile $\tilde{T}(t)$ is independent of the transverse coordinate $\overrightarrow{r}$ and $F$ is proportional to $|\tilde{E}_{\perp}|^{2}$. Consequently, distributions of $F^{1/2}(\overrightarrow{r},z)$ can be used to approximate $|\tilde{E}_{\perp}(\overrightarrow{r},z)|$ at a given position in vacuum $z$, and are used as inputs for the GSA-MD.
\textcolor{black}{For the LLC use case, the input dataset is a set of} 4 images of the laser fluence at positions $z=z_k$ spaced by $500$ \si{\micro\metre}.
Each image is the average of single-shot fluence measurements made at the corresponding plane $z_k$. In order to correct for small shot-to-shot pointing fluctuations, every individual measurement at a plane $z_k$ is centered on the transverse coordinates of its maximum before computation of the total average.

The \textcolor{black}{GSA-MD} procedure was done with \textcolor{black}{an HG basis, using} $N_m=N_n=30$ modes in each transverse direction, and numerical waists of $w_{0,x,y}=30$ \si{\micro\metre}. \textcolor{black}{The motivations behind the choice of these parameters are discussed in Appendix\,\ref{sec:appendix_GSAMD}.}

\begin{figure*}[ht!]
	\centering
	  \includegraphics[width=0.65\textwidth]{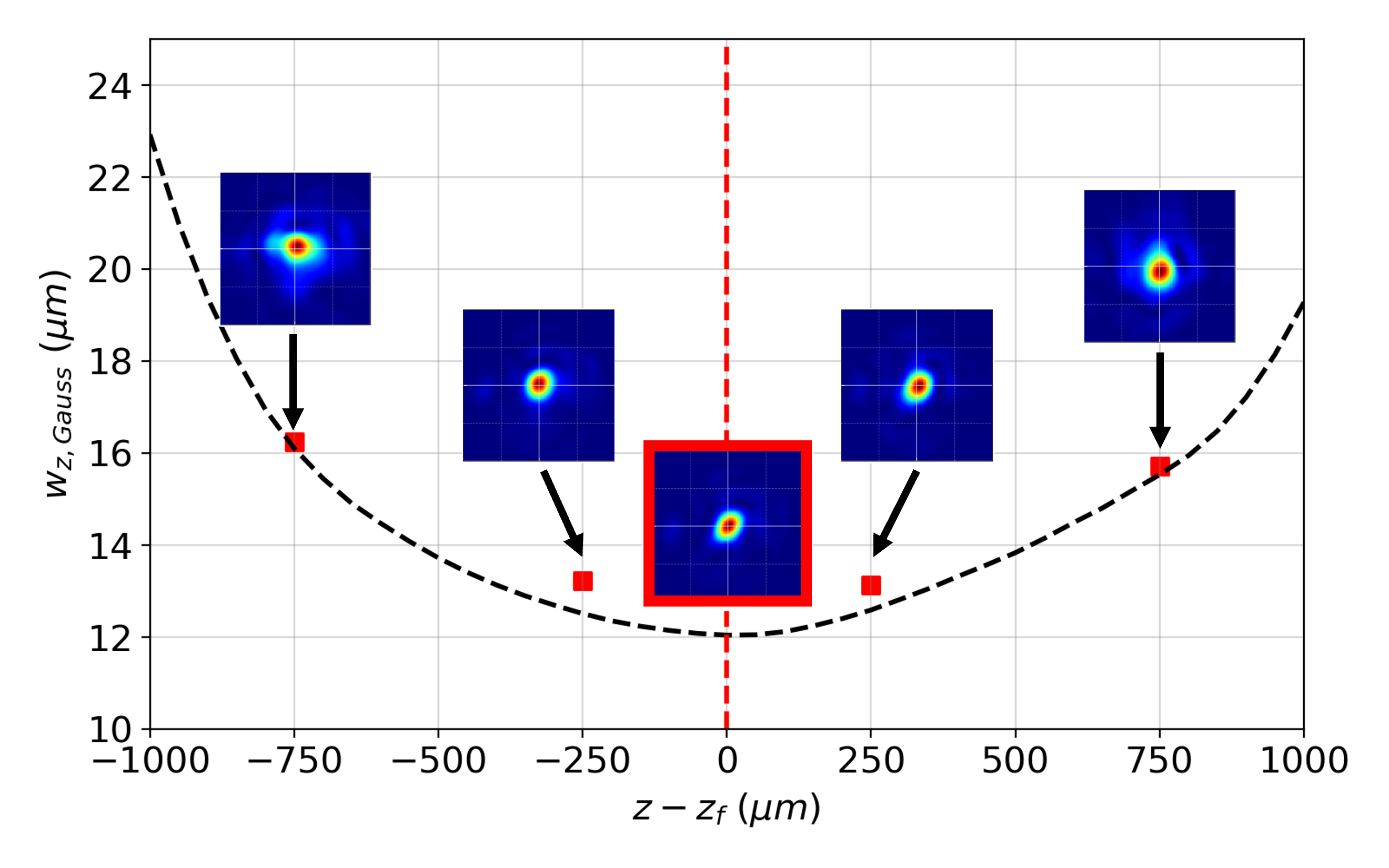}
	\caption{Evolution of $w_{z, Gauss}$, the laser waist determined by a Gaussian fit of the measured laser fluence distribution at each transverse plane, as a function of $z-z_f$, the laser relative position along the propagation axis.
    Red squares correspond to waist values determined on laser fluence images, while the dashed curve corresponds to the values predicted by the GSA-MD reconstruction. 
    The dashed vertical line corresponding to $z=z_f$ is the focal plane position predicted by the GSA-MD reconstruction. The fluence image in this plane, represented with a red border, was generated using the GSA-MD coefficients.}
	\label{fig:Waist_Lund}
\end{figure*}

\textcolor{black}{The output of the GSA-MD with HG basis is the set $\set{\tilde{C}_{m, n} \given m\in [0,\,1...,\,N_m],\,n\in [0,\,1...,\,N_n]}$ that minimizes the error between the fluence of the HG modal decomposition and experimental fluence images at positions $z=z_k$ of the laser propagation in vacuum~\cite{moulanier2023fast, massimo2025laser}.}

\textcolor{black}{Reconstructed }fluence images are shown along the laser propagation axis in Fig.\,\ref{fig:Waist_Lund}. Physical waist values, estimated by a radial Gaussian fit, are calculated for experimental fluence images (red squares). The reconstructed distribution (black dashed curve) is evaluated by propagating the HG modes obtained from the GSA-MD to any intermediate plane, i.e., where no experimental data is available. \textcolor{black}{The reconstruction} is in good agreement with the experimental data at the 4 planes used for GSA-MD ($z_k=-750,-250,+250,+750$ \si{\micro\metre}). The maximum difference between the reconstructed and the measured Gaussian waist is $0.5$ \si{\micro\metre}, which represents a relative error of $4$ \si{\percent}.

\subsection{Implementation in a PIC code}\label{sec:Implementation}

\textcolor{black}{The electric field is modelled as the product of user-defined transverse and longitudinal distributions. The temporal profile is calculated from the unchirped Gaussian profile of the LLC laser with $\tau=33.3$ \si{\femto\second}. 
The chirped longitudinal profile is then computed following the steps and sampling parameter of Sec.\,\ref{sec:Temporal_reconstruction}.
}

\textcolor{black}{To model the laser transverse distribution in the quasi-3D cylindrical geometry~\cite{lifschitz2009particle},} the original GSA-MD distribution obtained in an HG basis of transverse coordinates $(r_{0}, r_{1})=(x,y)$ and mode indexes $(k_{0}, k_{1})= (m, n)$ was converted to the LG basis of transverse coordinates $(r_{0^'}, r_{1^'})=(r, \theta)$ and mode indexes $(k_{0^'}, k_{1^'})= (l, p)$. \textcolor{black}{This conversion, referred to as ``Basis Swap" in Fig.\,\ref{fig:Framework}, was performed by projecting the GSA-MD electric field reconstructed in its focal plane over a basis of normalized LG modes.} \textcolor{black}{Appendix\,\ref{sec:appendix_transverse_distribution} details the steps of the basis swap and the assessment of its accuracy with respect to the original GSA-MD transverse distribution.}

\textcolor{black}{
For any truncated modal decomposition, i.e. with finite values of $N_{0,1}$ in Eq.\,(\ref{eq:Efield_spatial}), the corresponding distribution is not a perfect reconstruction of the measured laser. To quantify this difference, we define the energy ratio $\alpha$ between the two integrated distributions:
\begin{equation}\label{eq:a0_ratio}
    \alpha = \frac{\iint |{\tilde{\mathrm{e}}}_{\perp}(r_0, r_1, z=z^*)|^2\mathrm{det}[\mathbf{J}(r_0,r_1)] \mathop{dr_0} \mathop{dr_1}}{\iint f(r_0, r_1, z=z^*)\mathrm{det}[\mathbf{J}(r_0,r_1)] \mathop{dr_0} \mathop{dr_1}},
\end{equation}
where $z^*$ is a plane where the fluence has been measured close to focus ($z^*-z_f=-250$ \si{\micro\metre} in Fig.\,\ref{fig:Waist_Lund}), $f$ the experimental laser fluence distribution [Eq.\,(\ref{eq:fluence})] divided by its maximum, ${\tilde{\mathrm{e}}}_{\perp}$ the reconstructed electric field transverse distribution [Eq.\,(\ref{eq:Efield_spatial})] divided by its maximum. The quantity $\alpha$ represents the fraction of the laser transverse distribution that is effectively reconstructed by the distribution of $\tilde{E}_{\perp}$. It is also the fraction of the laser energy in the modelled distribution set to the same peak intensity as the measured one:
\begin{equation}\label{eq:E_ratio}
    E_{laser,sim} = \alpha \times E_{laser},
\end{equation}
where $E_{laser,sim}$ is the energy of the modelled distribution in ACE toolbox simulations (step labelled ``Energy Norm" in Fig.\,\ref{fig:Framework}). From Eqs.\,(\ref{eq:fluence}) and (\ref{eq:energy}), the energy scales with the squared amplitude of the electric field ($E_{laser}\propto a_0^2$). Consequently, in the ACE toolbox modelling, the electric field amplitude is scaled down by a factor $\sqrt{\alpha}$ compared to the value that would be required for the modelled laser energy to be $E_{laser}$.}
In the LG modelling of LLC transverse distribution, the value of $\alpha=85$\,\si{\percent} was measured at $z^*-z_{f}=-250$ \si{\micro\metre}.

\section{Results}\label{sec:results}

\subsection{Modelled laser distributions}\label{sec:sim_inputs}

\begin{figure}[h]
	\centering
	  \includegraphics[width=1.0\linewidth]{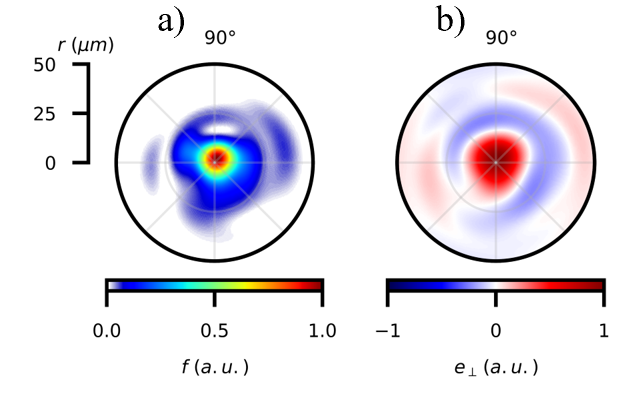}
	\caption{
	Transverse Laguerre-Gauss distribution of the laser modelled at the start of FBPIC simulations, i.e. $z-z_{cell}=-1300$ \si{\micro\metre} in the simulation propagation or $z-z_{f}=-550$ \si{\micro\metre} on the vacuum propagation axis in Fig.\,\ref{fig:Waist_Lund}: a) $f$, the laser fluence divided by its maximum – b) $e_{\perp}$, the transverse part of the electric field at the co-moving coordinate of the peak intensity, divided by its maximum.
    }
	\label{fig:Transverse_distribution}
\end{figure}

\begin{figure}[h]
	\centering
	  \includegraphics[width=1.0\linewidth]{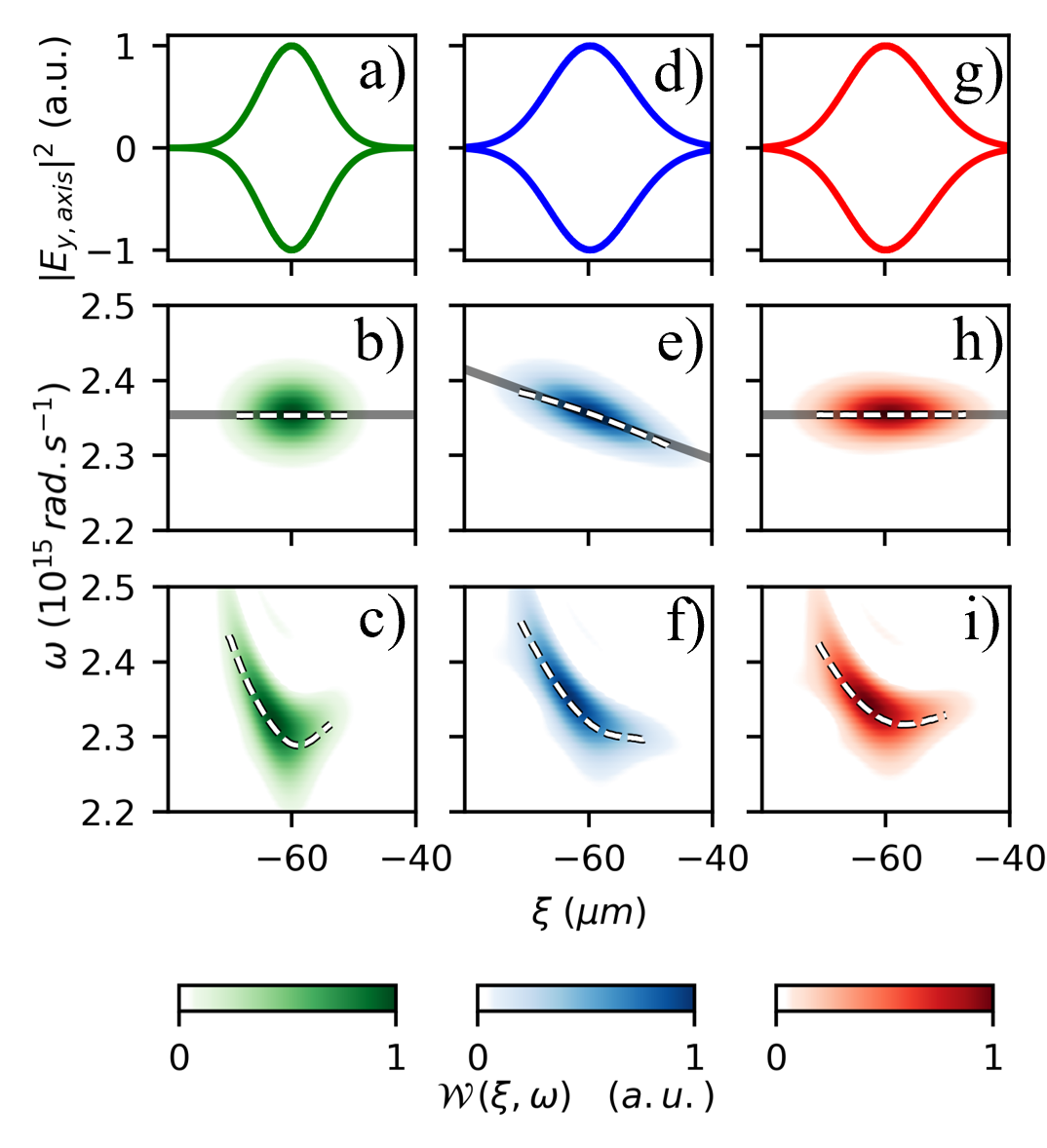}
	\caption{Laser temporal envelope for 3 PIC simulation cases which chirp parameters are specified in Table \ref{tab:Lund_chirped_sims_parameters}:
    a) - c) TGS (green) - d) - f) TCS (blue)- g) - i) TAS (red).
    The first row shows $|E_{y,axis}|^2$, the squared envelope of the electric field on the propagation axis and obtained by Hilbert transform divided by its maximum, and its negative $-|E_{y,axis}|^2$, at $z-z_{cell}=-1300$ \si{\micro\metre} (start of simulation). The second and third rows show the normalized Wigner transform of $E_{y,axis}$ at $z-z_{cell}=-1300$ and $z-z_{cell}=300$ \si{\micro\metre} (middle of electron injection) respectively. On the second and third rows, the dashed curve shows $\omega(\xi)$, the average frequencies of the Wigner distribution along the $\xi$-axis, and the black curve in the second row is the theoretical solution of Eq.\,(\ref{eq:omega_chirp}) using the input $\phi_2$ value of each simulation.
    }
	\label{fig:Chirp_Lund}
\end{figure}

\textcolor{black}{The shared numerical parameters of the simulation grid, laser transverse distribution and target plasma density profile are summarized in Table\,\ref{tab:Lund_sims_parameters} and discussed in Appendix\,\ref{sec:appendix_sims}.
The simulated Laguerre-Gauss transverse distribution used for all simulations is represented in Fig.\,\ref{fig:Transverse_distribution}, at $z-z_{cell}=-1300$ \si{\micro\metre}, where $z_{cell}=1300$ \si{\micro\metre} is the position of the gas cell entrance in simulation coordinates.}

\setlength{\tabcolsep}{18pt}
\begin{table}[ht!]
\centering
\normalsize
    \begin{tabular}{cc}
    \toprule[.1em]
    \addlinespace
    $\Delta z/\lambda_0$ & $1/32$ \\
    $\Delta r/\lambda_0$ & $1/2$ \\
    $\left[P_z,\,P_r,\,P_{\theta}\right]$ & $\left[2,\,2,\,16\right]$ \\
    $\gamma_b$ & 1.5 \\
    \addlinespace 
    \hline
    \addlinespace 
    $E_{laser, sim}$ ($\si{J}$) & 0.74 \\
    $w_{0,LG}$ (\si{\micro\m}) & $30$ \\
    $\left[N_l,\,N_p\right]$ & $\left[3,\,30\right]$ \\
    $N_\theta$ & $4$ \\
    \addlinespace 
    \hline
    \addlinespace
    $n_0$ (\si{\per\cm\tothe{3}}) & $4.5\times\si{10\tothe{18}}$ \\
    $n_{\mathrm{N}} / n_{\mathrm{H}}$ & $5\,\%$\\
    \addlinespace
    \bottomrule[.1em]
    \end{tabular}
    \caption{
    \textcolor{black}{Main parameters shared across FBPIC simulations presented in Section\,\ref{sec:results}:
    $\Delta z/\lambda_{0}$ and $\Delta r/\lambda_{0}$, longitudinal and radial step sizes of the simulation grid; $\left[P_z,\,P_r,\,P_{\theta}\right]$, the number of macro-particles per cell along $z$ and $r$, and per $2\pi$ along $\theta$; $\gamma_b$, the Lorentz factor of the boosted frame; $E_{laser,sim}$, the modelled ACE laser energy; $w_{0,LG}$, the numerical waist of each LG mode; $\left[N_l,\,N_p\right]$, the maximum azimuthal and radial orders of the laser transverse LG decomposition; $N_\theta$, maximum harmonic order used for the Maxwell's Equations solver; $n_0$, the peak plasma electron density and $n_{\mathrm{N}} / n_{\mathrm{H}}$, the percentage of Nitrogen dopant for ionization injection.}}
    \label{tab:Lund_sims_parameters}
\end{table}

\setlength{\tabcolsep}{12pt}
\begin{table}[ht!]
\centering
\normalsize
    \begin{tabular}{cccc}
    \toprule[.1em]
    \addlinespace 
    \textbf{Parameter} & \textbf{TGS} & \textbf{TCS} & \textbf{TAS}\\
    \addlinespace 
    \toprule[.1em]
    \addlinespace 
    $\phi_2$ (\si{\femto\second\squared}) & $0$ & $+501$ & $0$\\
    $\phi_3$ (\si{\femto\second\cubed})  & $0$ & $-3317$ & $0$\\
    $\phi_4$ (\si{\femto\second\tothe{4}}) & $0$ & $+4485$ & $0$\\
    \addlinespace 
    \hline
    \addlinespace 
    $\chi^{-1/4}$ & $1$ & $0.85$ & $0.85$\\
    $\tau_{b}$ (\si{\femto\second}) & $33.3$ & $42.2$ & $42.2$\\
    $\tau_{f}$ (\si{\femto\second}) & $33.3$ & $47.7$ & $47.7$\\
    \bottomrule[.1em]
    \end{tabular}
    \caption{
    Parameters of temporal profiles used for FBPIC simulations with an unchirped electric field (``Temporally Gaussian Simulation" or TGS), 3D reconstructed chirped electric field distribution (``Temporally Chirped Simulation" or TCS) and electric field with a temporal bi-Gaussian fit of the asymmetric chirped profile (``Temporally Asymmetric Simulation" or TAS).
    For each simulation, the following parameters are specified:
    chirp coefficients from second to fourth order, $\phi_{2}$, $\phi_3$, $\phi_4$, chirp reduction factor of the laser electric field amplitude $\chi^{-1/4}$,
    laser $1/e^2$ duration at the back of its temporal profile, $\tau_{b}$,
    laser $1/e^2$ duration at the front of its temporal profile, $\tau_{f}$.
    }
    \label{tab:Lund_chirped_sims_parameters}
\end{table}

The effects of the laser temporal profile chirping are studied by comparing three different simulations with the same transverse parameters and different temporal profiles which parameters are listed in Table\,\ref{tab:Lund_chirped_sims_parameters}. The ``Temporally Gaussian Simulation" (TGS) is a simulation with an unchirped Gaussian profile ($\phi_2=\phi_3=\phi_4=0$). The effects of a chirped temporal profile are modelled in two types of simulations : firstly, the ``Temporally Chirped Simulation" (TCS), in which the chirped temporal profile is modelled using the steps described in section~\ref{sec:Temporal_reconstruction} and Fig.\,\ref{fig:Framework} ``Temporal" block, with $\phi_2=501$ \si{\femto\second\squared}, $\phi_3=-3317$ \si{\femto\second\tothe{3}}, $\phi_4=4485$ \si{\femto\second\tothe{4}} - secondly the ``Temporally Asymmetric Simulation" (TAS), in which the shape of the temporally chirped profile is reconstructed by a bi-Gaussian profile with a front duration $\tau_f=47.7$ \si{\femto\second}, back duration $\tau_b=42.2$ \si{\femto\second}, with no spectral phase shaping (i.e. $\phi_2=\phi_3=\phi_4=0$) and peak amplitude scaled down by a factor $\chi^{-1/4}=0.85$ calculated from Eq.\,(\ref{eq:a_chirp}) with the $\phi_2$ value of the TCS.
The TCS was used in a previous study of the LLC experiment~\cite{Backhouse2025}, in which the accuracy of the ACE toolbox was shown by the agreement between the energy-angle spectrum in the TCS and multiple experimental spectra.

The temporal profiles at the start of simulation ($z-z_{cell}=-1300$ \si{\micro\metre}) are plotted in the first row of Fig.\,\ref{fig:Chirp_Lund} for all three cases. The second and third rows represent the distribution of frequencies at the start of simulation and during the electron bunch injection ($z-z_{cell}=300$ \si{\micro\metre}). This distribution is calculated with the Wigner transform of the transverse laser electric field on the propagation axis~\cite{paye1995space, hong2002time}:
\begin{equation}\label{eq:Wigner_transform}
    \begin{aligned}
        &\mathcal{W}(\xi, k)=\frac{1}{\pi}\int_{-\infty}^{+\infty} E_{\perp,\xi}\left(\xi-\frac{\xi'}{2}\right) E_{\perp,\xi}^*\left(\xi+\frac{\xi'}{2}\right) \nonumber \\
        &\hspace{90pt} \times \exp\left[-ik \xi'\right] \mathop{d\xi'},
    \end{aligned}
\end{equation}
where $E_{\perp,\xi}$ is the on-axis real part of the electric field (at $r=0$) and $k$ the wave number.

The initial temporal profiles in the TGS and TAS [Figs.\,\ref{fig:Chirp_Lund}.(a), (b) and (g), (h)] have a flat distribution of instantaneous frequencies. This is shown by the calculation of $\omega(\xi) \simeq \langle k(\xi) \rangle c$ (dashed white curves), where $\langle . \rangle$ is the average along the $\xi$-axis over the distribution of $\mathcal{W}(\xi,k)$~\cite{pathak2012effect}. For the TGS and TAS, the horizontal curve $\langle \omega(\xi) \rangle$ is positioned as expected at the carrier frequency $\omega(\xi) \simeq \omega_0 = 2\pi c/ \lambda_0$. In comparison, the initial temporal distribution of instantaneous frequencies in vacuum for the TCS [Fig.\,\ref{fig:Chirp_Lund}.(d), (e)] is linear with a negative slope. This is in agreement with the theoretical Gaussian laser chirped with a second order coefficient~\cite{hong2002time}:
\begin{equation}\label{eq:omega_chirp}
    \omega(\xi) = \omega_0 - \dfrac{4\phi_2}{\left( \tau_{\mathrm{FWHM}}^4 + 4\phi_2^2 \right)}\,\dfrac{\left(\xi - \xi_0\right)}{c} \,,
\end{equation}
 where $\tau_{\mathrm{FWHM}} = \tau_{l}\sqrt{2\ln{\left(2\right)}}$ is the full-width at half-maximum duration of the unchirped temporal profile.

\subsection{Comparison of electron characteristics}\label{sec:dynamics}

\begin{figure}[h]
	\centering
	  \includegraphics[width=\linewidth]{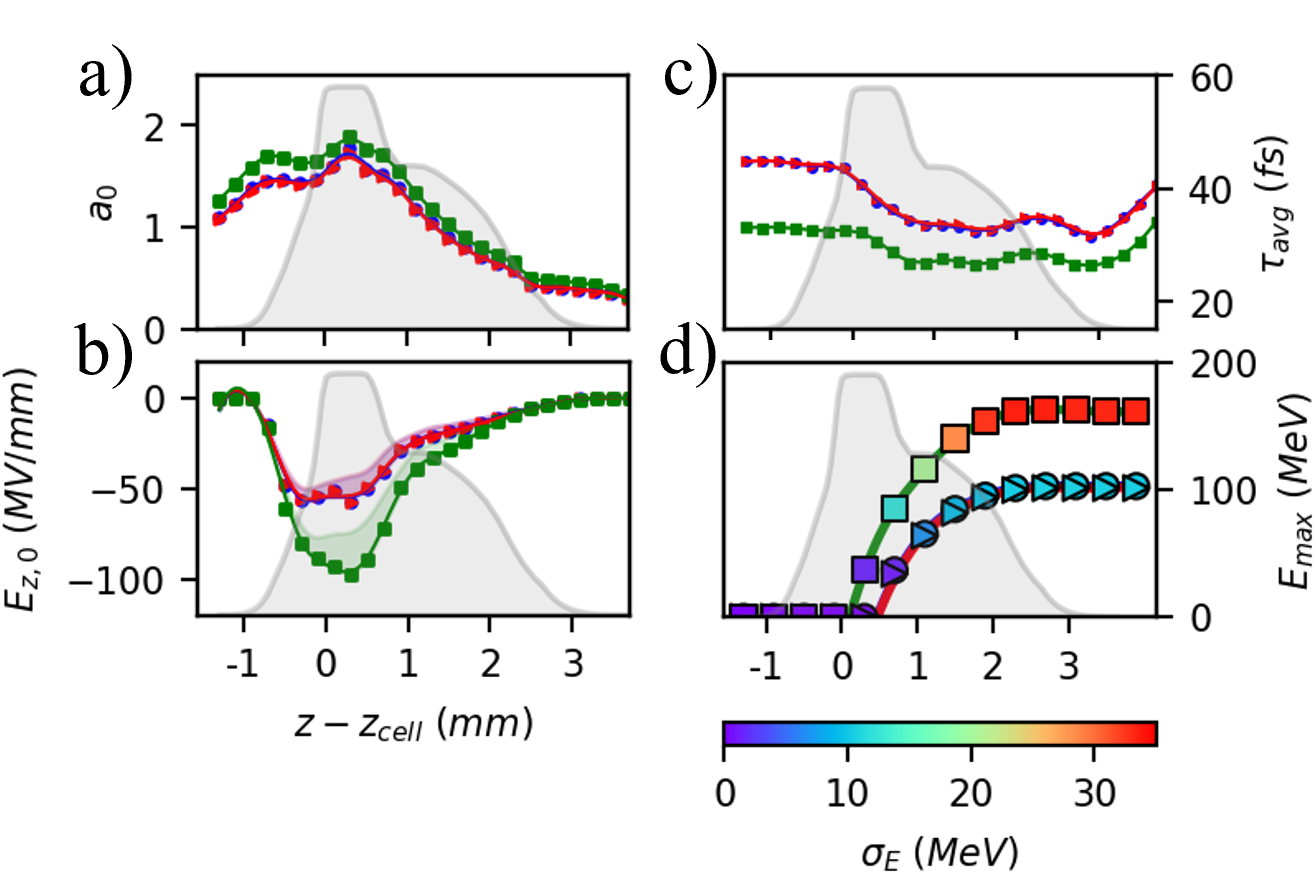}
	\caption{
	Evolution of LWFA parameters for the 3 laser configurations: TGS (green curves and squares), TCS (blue curves and circles) and TAS (red curves and triangles).
    a) Maximum normalized laser amplitude $a_0$ - b) electric field amplitude $E_{z,0}$ (solid curves), and associated standard deviation (shaded areas) in a $5$ \si{\micro\metre} window - c) average $1/e^2$ duration $\tau_{avg} = (\tau_{b}+\tau_{f})/2$ - d) $E_{max}$, maximum energy for which $dQ/dE \geq 0$ \si{pC/MeV}, and in colorscale the different values of $\sigma_E$, standard deviation of the electrons accelerated in the first plasma cavity.
    In each plot, $n_e$, the plasma density profile, is represented by the gray shaded area.
    }
	\label{fig:Lund_sims_comparison}
\end{figure}

The laser temporal profiles for the TGS, TCS and TAS, associated to the same reconstructed transverse field distribution, were used for three simulations of LWFA with ionization injection. For these simulations, the peak electron density was set at $n_0=4.5\times$\si{10\tothe{18}} \si{\cm\tothe{-3}}. The background plasma longitudinal density profile, taking into account ionization levels up to $\mathrm{H^+}$ and $\mathrm{N^{5+}}$, is represented by the gray shaded area in Fig.\,\ref{fig:Lund_sims_comparison}. The start of the plasma density plateau along the longitudinal axis is located at the simulation position $z=z_{cell}$ and spans over $500$ \si{\micro\metre}.

The evolutions of the laser and injected electrons are plotted along the longitudinal axis in Fig.\,\ref{fig:Lund_sims_comparison}: the laser peak normalized amplitude $a_0$ in Fig.\,\ref{fig:Lund_sims_comparison}.(a), the peak on-axis longitudinal electric field $E_{z,0}$ in Fig.\,\ref{fig:Lund_sims_comparison}.(b), the laser average duration $\tau_{avg}=(\tau_{b}+\tau_{f})/2$ in Fig.\,\ref{fig:Lund_sims_comparison}.(c), $E_{max}$, the maximum electron energy for which $dQ/dE>0$ \si{\pico\coulomb\per\mega\electronvolt} in Fig.\,\ref{fig:Lund_sims_comparison}.(d) right-hand axis, along with $\sigma_E$, the standard deviation of the macro-particles weighted by their spectral charge $dQ/dE$ in colored markers.

The TCS and TAS have similar laser and electron parameters throughout the propagation, with $a_0$ and $\tau_{avg}$ following the exact same curves.
This shows that the initial frequency chirp in Fig.\,\ref{fig:Chirp_Lund}.(e) plays a negligible role during the propagation in comparison to envelope shaping. In the modelled regime, this can be inferred by comparing the chirp term $\delta \omega/\omega_0$ (normalized instantaneous frequency perturbation) to the other contributions in the propagation equation that quantifies the evolution of the pulse duration~\cite{pathak2012effect}:

\begin{equation}\label{eq:tau_propagation}
    \dfrac{1}{c\tau_l}\dfrac{\partial\tau_l}{\partial t} = \dfrac{\omega_p^2}{2\omega_0^2}\dfrac{\partial}{\partial \xi}\left(- \dfrac{\delta n}{n_0} + \dfrac{\langle a^2 \rangle}{2} + 2\dfrac{\delta \omega}{\omega_0}\right) \,,
\end{equation}
where $\omega_p=\sqrt{n_0 \mathrm{e}^2/m_e\epsilon_0}$ is the plasma frequency, $\delta n$ the local plasma perturbation and $\langle a^2 \rangle$ the envelope of the squared laser normalized amplitude.
The importance of the frequency distribution is given by comparing its magnitude in Eq.\,(\ref{eq:tau_propagation}), $\mathrm{max}(|\partial_\xi (2\delta \omega/\omega_0)|)$, to that the gradients of the local density perturbation $\mathrm{max}(|\partial_\xi (\delta n/n_0)|)$ and laser envelope $\mathrm{max}(|\partial_\xi (\langle a^2 \rangle/2)|)$. In the TCS, the amplitudes of the local perturbation and laser envelope terms were estimated at the position $z-z_{cell}=300$ \si{\micro\metre} where $a_0$ is maximum. The gradient of the chirp term was calculated in the same plane and over the distribution of $\omega(\xi)$ plotted in Fig.\,\ref{fig:Chirp_Lund}.(f).
\textcolor{black}{The calculations of the different contributions yield $ \mathrm{max}(|\partial_\xi (2\delta \omega/\omega_0)|)\simeq 0.014 \ll \mathrm{max}(|\partial_\xi (\langle a^2 \rangle/2)|)\simeq 0.11 < \mathrm{max}(|\partial_\xi (\delta n/n_0)|)\simeq 0.30$ \si{\per\micro\metre}. }
This ordering confirms the minor role of the initial temporal frequency distribution in the laser evolution.
Furthermore, the contribution of the initial chirp coefficients [Fig.\,\ref{fig:Chirp_Lund}.(e)] is even more negligible compared to the chirp accumulated during propagation: the initial linear slope of frequencies corresponds to \textcolor{black}{$|\partial_\xi (2\delta \omega/\omega_0)|= 2.5\times\si{10\tothe{-3}}$ \si{\per\micro\metre}.}

\subsection{Parametric study of the second order chirp coefficient}

In this section, we study the effect of $\phi_2$ on the shaping of the spectral charge distribution. For $\phi_3=-3317$ \si{\femto\second\cubed} and $\phi_4=4485$ \si{\femto\second\tothe{4}}, 11 simulations were performed using $\phi_2$ values ranging from $0$ to $501$ \si{\femto\second\squared}. The explored values were spaced by $50$ \si{\femto\second\squared}, up to the 2 last points spaced by $51$ \si{\femto\second\squared}.

\begin{figure}[h]
	\centering
	  \includegraphics[width=1.0\linewidth]{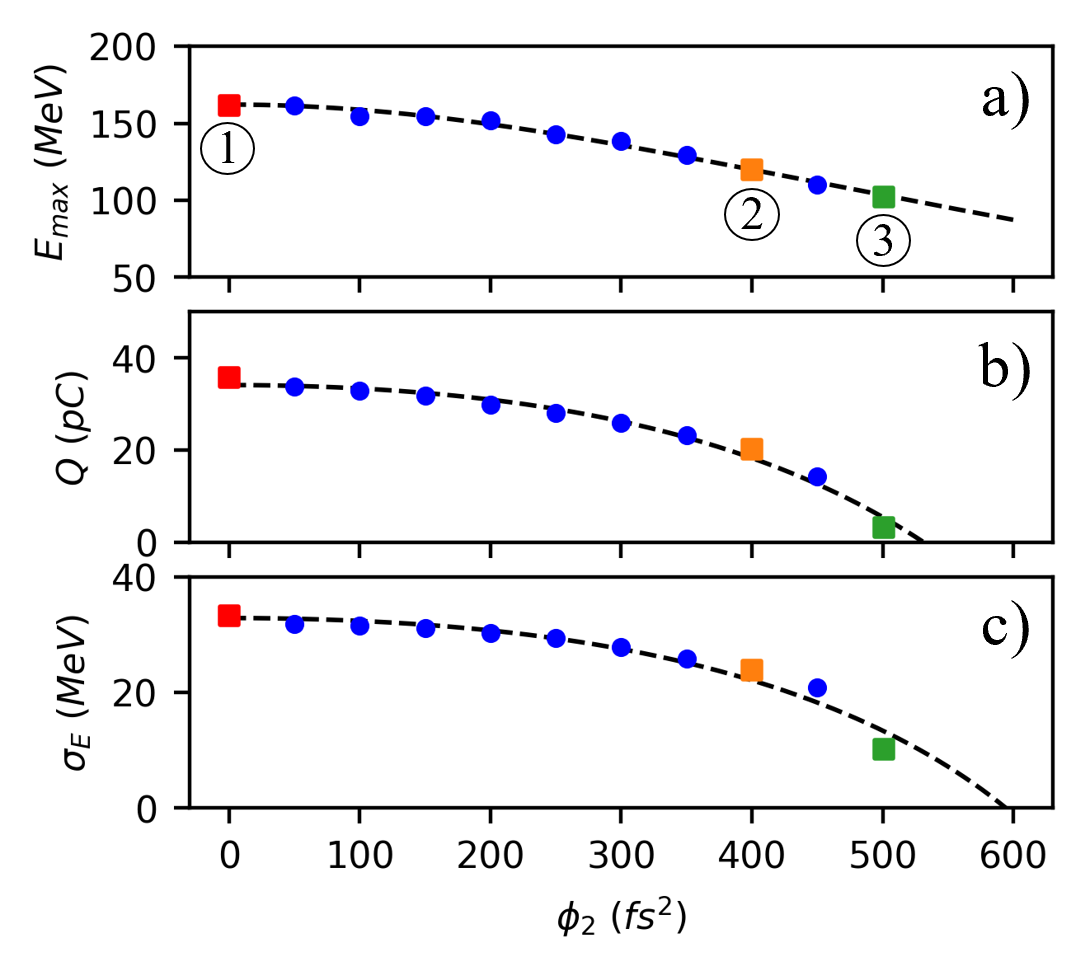}
	\caption{Electron parameters at the end of simulation as a function of the second order chirp $\phi_2$ and for fixed values of $\phi_3=-3317$ \si{\femto\second\tothe{3}} and $\phi_4=4485$ \si{fs^4}: a) $E_{max}$, highest energy for which $dQ/dE > 0$ \si{\pico\coulomb\per\mega\electronvolt} - b) $Q$, charge of the electrons accelerated in the first plasma cavity - c) $\sigma_E$, standard deviation of the electrons accelerated in the first plasma cavity. In all plots, the red square is the point for which $\phi_2=0$ \si{\femto\second\tothe{2}} (labelled 1), the orange square $\phi_2=400$ \si{\femto\second\tothe{2}} (labelled 2) and green square $\phi_2=501$ \si{\femto\second\tothe{2}} (labelled 3), which corresponds to the TCS case. $dQ/dE$ profiles for these three points are plotted in Fig.\,\ref{fig:chirp_dQdE}.
	}
	\label{fig:chirp_scan}
\end{figure}

The main parameters characterizing the energy spectrum of the accelerated electrons, which are the bunch maximum energy $E_{max}$, charge $Q$, and standard deviation $\sigma_E$, are plotted in Figs.\,\ref{fig:chirp_scan}.(a)-(c).

The maximum energy $E_{max}$ is an indicator of the bunch energy gain and depends on the amplitude of the longitudinal electric field of the plasma bucket in which the electrons are injected. This amplitude was tracked in simulations, as shown in Fig.\,\ref{fig:Lund_sims_comparison}.(b).
The dependence of $E_{max}$ with $\phi_2$ is explained by the variation of the amplitude of the accelerating electric field with the decrease of $a_0$. In the 3D nonlinear regime for which $a_0>2$, $E_{max}$ is expected to scale with $a_0$~\cite{lu2007generating, couperus2017demonstration}. Since the values of $a_0$ in the TGS and TCS peak to $1.9$ and $1.7$ respectively [Fig.\,\ref{fig:Lund_sims_comparison}.(a)], a linear relation between $E_{max}$ and $a_{0,c}(\phi_2)$ [Eq.\,(\ref{eq:E_fit})] was used in Fig.\,\ref{fig:chirp_scan}.(a) to fit $E_{max}$ as a function of $\phi_2$. The details of the fit are described in Appendix\,\ref{sec:appendix_fit_E}.

As described in Appendix\,\ref{sec:appendix_fit_sigmaE}, the evolution of $Q$ with $\phi_2$ was fitted as a function of the volume of the laser distribution inside which electrons of $\mathrm{N^{5+}}$ can be ionized. Since the transverse distribution of this volume remains the same when changing $\phi_2$ in the absence of STC, the fit was expressed as a function of $a_{0,c}$, the chirped laser amplitude, and $\tau_{l,c}$, the chirped $1/e^2$ duration. 

Figs.\,\ref{fig:chirp_scan}.(b) and (c) show that the energy dispersion and bunch charge have the same behavior and can be fitted with the same function [Eq.\,(\ref{eq:sigmaE_fit})] and different constants
Both fits predict that, for $\phi_3=-3317$ \si{\femto\second\cubed} and $\phi_4=4485$ \si{\femto\second\tothe{4}}, no bunch would be injected for values of $\phi_2\sim600$ \si{\femto\second\squared}.

\begin{figure}[h]
	\centering
	  \includegraphics[width=1.0\linewidth]{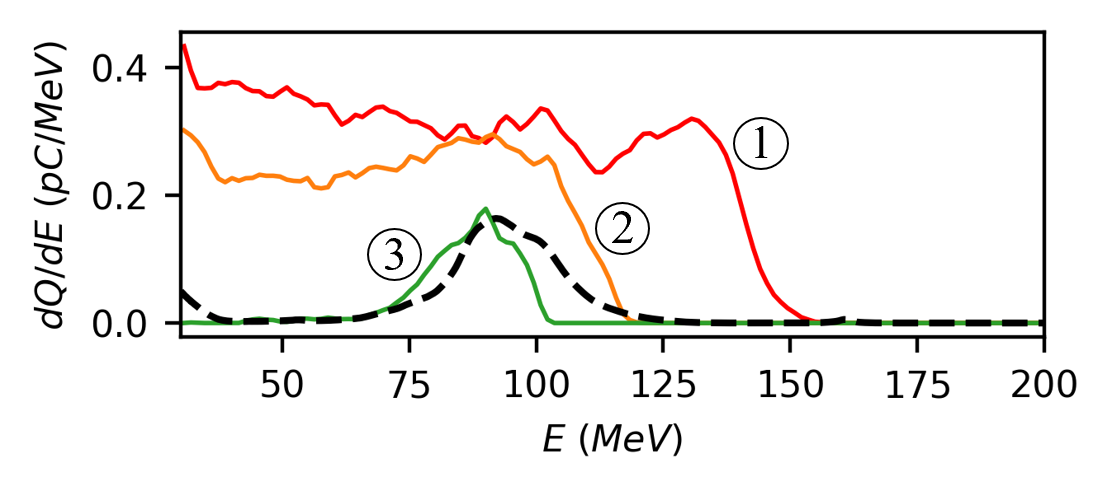}
	\caption{Spectral charge $dQ/dE$ at the end of simulation ($z - z_{cell}=3700$ \si{\micro\metre}) as a function of energy for fixed values $\phi_3=-3317$ \si{\femto\second\cubed} and $\phi_4=4485$ \si{\femto\second\tothe{4}}, and with 3 values of $\phi_2$: $\phi_2=0$ \si{\femto\second\squared} (red solid curve, labelled 1), $\phi_2 = 400$ \si{\femto\second\squared} (yellow solid curve, labelled 2), $\phi_2 = 501$ \si{\femto\second\squared} (green solid curve, labelled 3). The average $dQ/dE$ profile of 10 consecutive shots produced with the experimental chirp coefficients $\phi_2 = 501$ \si{\femto\second\squared}, $\phi_3=-3317$ \si{\femto\second\cubed} and $\phi_4=4485$ \si{\femto\second\tothe{4}} is represented by the black dashed curve.
    Each labelled curve is associated to the points of matching label and color in Figs.\,\ref{fig:chirp_scan}.(a), (b) and (c).}
	\label{fig:chirp_dQdE}
\end{figure}

The effects of the modulations of $\phi_2$ on electron spectra are shown in Fig.\,\ref{fig:chirp_dQdE}, where $dQ/dE$ is plotted at the end of simulations as a function of $E$ for $\phi_2=0$ \si{\femto\second\squared} (simulation labelled 1), $\phi_2=400$ \si{\femto\second\squared} (labelled 2) and $\phi_2=501$ \si{\femto\second\squared} (labelled 3). The increase in $\phi_2$ leads to the emergence of a peak in the spectrum \textcolor{black}{due to the trapping and acceleration of a smaller fraction of electrons created by ionization injection.}
In Fig.\,\ref{fig:chirp_dQdE}, starting from a broad profile at $\phi_2=0$ \si{\femto\second\squared}, a single peak arises from $\phi_2=400$ \si{\femto\second\squared}, which corresponds to the elbow of the data points for $Q$ and $\sigma_E$ in Figs.\,\ref{fig:chirp_scan}.(b) and (c). 
For the TCS value $\phi_2=501$ \si{\femto\second\squared}, the simulated single peak is in good agreement with the average profile of 10 consecutive shots in the experiment, for which $\phi_2=501$ \si{\femto\second\squared}, $\phi_3=-3317$ \si{\femto\second\cubed} and $\phi_4=4485$ \si{\femto\second\tothe{4}}. 
This correspondence is highlighted by the energy of the peaks, with $90$ \si{\mega\electronvolt} in simulation against $92$ \si{\mega\electronvolt} in the experiment, and peak spectral charge, with $0.18$ \si{\pico\coulomb\per\mega\electronvolt} against $0.16$ \si{\pico\coulomb\per\mega\electronvolt}, and validates the methods of the ACE toolbox for the modelling of the use case experiment.

\section{Conclusions}

We described the ACE toolbox, which can be used to process experimental data for the 3D modelling of an electric field with a chirped temporal profile and an imperfect asymmetrical transverse distribution. The ACE toolbox is composed of modules able to model experimental data and use them as input for PIC simulations. Under the assumption of no spatio-temporal couplings, the temporal and transverse reconstruction are completely independent from one another.

The quasi-3D simulations initialized with the ACE reconstructed laser give a physical insight of the mechanisms leading to the low energy spread working point found in the experiment by varying the chirp coefficients. In this regime, where the laser amplitude is just above the ionization injection threshold, fine tuning of the temporal envelope shape through control of its flatness (second order chirp) and asymmetry (third order chirp) demonstrated an equivalent control over charge, maximum energy and energy spread. The second order in particular has the biggest impact on the energy spread through ionization, as demonstrated by the parametric study performed \textcolor{black}{in simulations}. The effect of the third order chirp coefficient on the bunch injection was studied in~\cite{Backhouse2025}. Simulation results show a dependence of the wakefield amplitude with the skewness of the laser temporal profile controlled by the sign of $\phi_3$. 

The accuracy of the ACE toolbox and its modularity makes it suitable for low to very high resolution PIC simulations of LWFA with a chirped laser. This makes the presented tools compatible with multi-objective numerical Bayesian optimization schemes that rely on a large number of input parameters~\cite{irshad2023multi, irshad2024pareto}, including spectral shaping~\cite{djordjevic2025bayesian}.

\section{Acknowledgments}
I. M. was supported by the CNRS in the framework of the project DIANA, Contract No. 1255841 and the EUPRAXIA Preparatory Phase (PP) Project, Contract No. 101079773.

The authors acknowledge M. Bisson for the design and management of the MAITRO HPC cluster at LPGP, providing computing resources for data analysis, software development and simulations for the work presented on this article.

Experimental data used in this article were obtained during an experiment conducted at the Lund Laser Centre in 2021 which received funding from the European Union’s Horizon 2020 Research and Innovation Programme under Grant Agreement No. 730871. We thank the following collaborators for providing these data: M.~P.~Backhouse, L.~T.~Dickson, C.~C.~Cobo, F.~Filippi, C.~Gustafsson, E.~Lofquist, K.~Svendsen, M.~J.~V.~Streeter, R.~J.~Shalloo, O.~Vasilovici, C.~Ballage, S.~J.~D.~Dann, C.~D.~Murphy, Z.~Najmudin, S.~Dobosz Dufrénoy and O.~Lundh. 

\vspace{0.5cm}

\appendix

\section*{Appendix}

\textcolor{black}{\subsection{GSA-MD parameters}\label{sec:appendix_GSAMD}}

\textcolor{black}{
In this appendix, the motivations behind the choice of the main parameters used for the ACE transverse reconstruction are discussed. Specifically, we explain how the transverse fluctuations and variations along the propagation axis of the reconstructed distribution are controlled by the numerical waist and higher orders of the mode basis.}

\textcolor{black}{
The fluence images recorded by a CCD camera in vacuum are defined over a Cartesian grid. This geometry prompted the use of the HG modes for the GSA-MD, which are defined in the Cartesian space.}

\textcolor{black}{
In the case of the HG basis, the main parameters impacting the GSA-MD performance are $N_m$ (maximum order in $x$) and $N_n$ (maximum order in $y$), and $w_{0,x}$, $w_{0,y}$, although they are usually set as $N_{m}=N_{n}=N$ and $w_{0,x}=w_{0,y}=w_0$. Within this Appendix, this case will be studied with the 1D profiles of HG mode intensities~\cite{Siegman86}:
\begin{equation}\label{eq:HGmodes}
    \begin{aligned}
        I_m(x,z)&=I_{0,m} \thinspace h^2_m\left[\sqrt{2}\frac{x}{w(z)}\right]\exp\left[-\frac{2x^2}{w^2(z)}\right]\\
        \frac{w(z)}{w_{0}}&=\sqrt{1+\left(\frac{z - z_f}{z_{R}}\right)^2}\\
        I_{0,m} &= \left(w(z)2^{m - 1/2}m!\sqrt{\pi}\right)^{-1},
    \end{aligned}
\end{equation}
where $I_m=|\mathrm{HG}_m|^2$, $h_m$ is the Hermite polynomial of order $m$ and $z_{R}=k_0w^2_{0}/2$ the Rayleigh length of an HG mode.
The user-defined pair \{$w_0$, $N$\} has an incidence on two characteristic scaling lengths: the radial resolution $\delta r$, and the radial extent $r_{E}$. At focus ($z=z_f$), we define the former as the minimum variation length that occurs within the oscillations of the set of intensities $\set{I_m(x,z_f) \given m\in [0,1,...,N]}$, and the latter as the maximum among radii containing $99$\,\si{\percent} of the integrals of $\set{I_m(x,z_f) \given m\in [0,1,...,N]}$.
}

\begin{figure}[h]
	\centering
	  \includegraphics[width=0.9\linewidth]{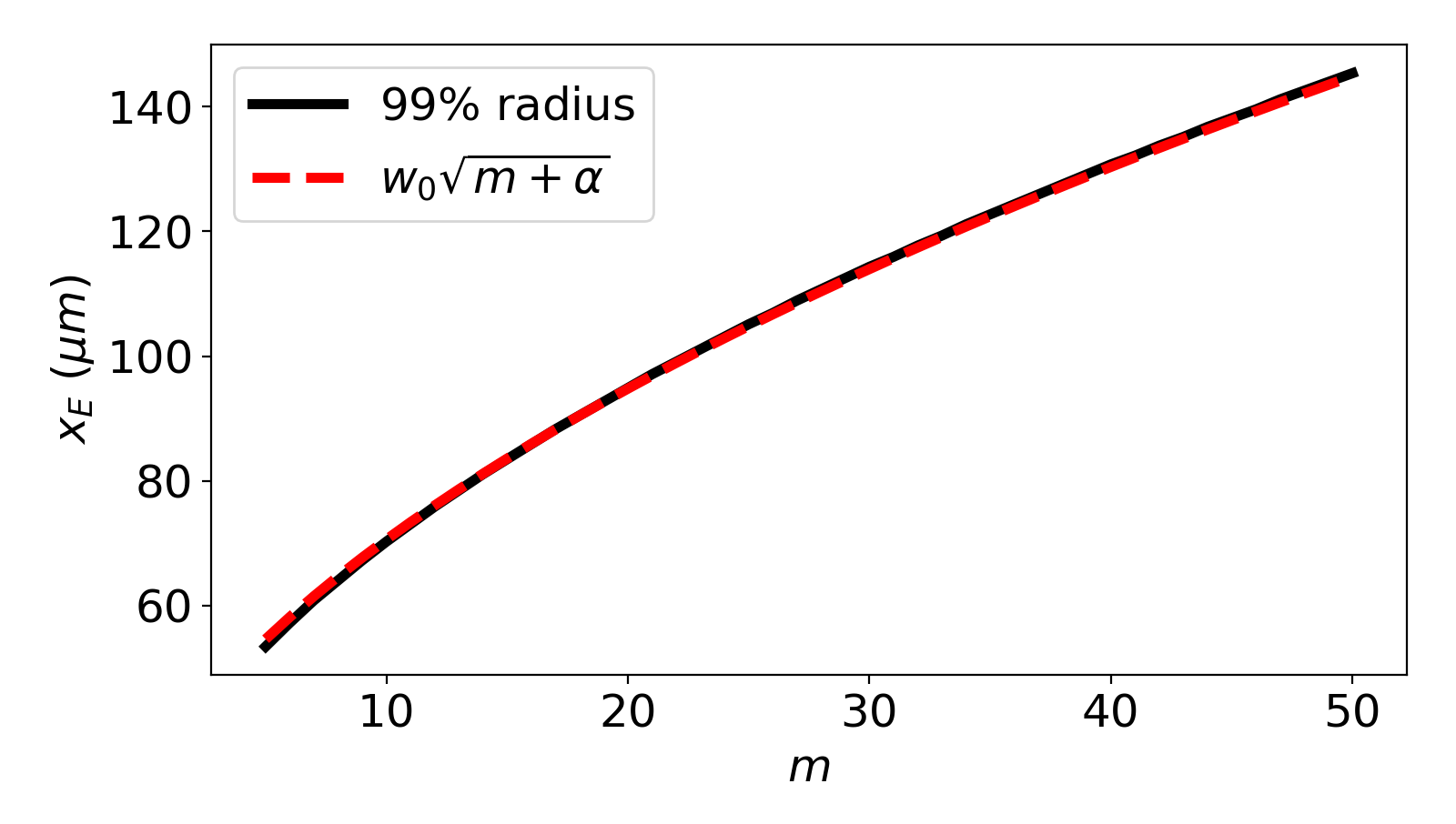}
	\caption{\textcolor{black}{Figure 3.2 from~\cite{moulanier2024modelisation}. Evolution of $x_E$, the radius containing $99$\,\si{\percent} of the integral of $I_m(x,z_f)$ as a function of $m$, its mode order, for $w_{0}=20$ \si{\micro\metre}. The black curve represents the calculated curve for $m$ between $5$ and $50$, and the red curve is a least square fit yielding $x_{E,fit}=w_0\sqrt{m + 2.46}$.}}
	\label{fig:Radius_law}
\end{figure}

\textcolor{black}{
Fig.\,\ref{fig:Radius_law} represents $x_{E}$, the $99$\,\si{\percent}-radius of $\int I_m(x,z_f) dx$, as a function of $m$. For $N \gg 1$, the asymptotic behaviour of $r_{E}$ is:
\begin{equation}
    r_{E} \propto w_0\sqrt{N}.
\end{equation}
This scaling gives an order of magnitude of the maximum extent of the mode basis.
The orthogonal condition of the mode basis defined in Eq.\,(\ref{eq:Efield_spatial_2}) is always verified in $\mathbb{R}^2$ as any HG mode is decreased sufficiently close to zero over $r_{E}$. However, within the discretized grid used for the GSA-MD, the orthogonality of the mode basis requires $r_{E}$ to be smaller than the rectangular grid boundaries. 
}

\begin{figure}[h]
	\centering
	  \includegraphics[width=0.9\linewidth]{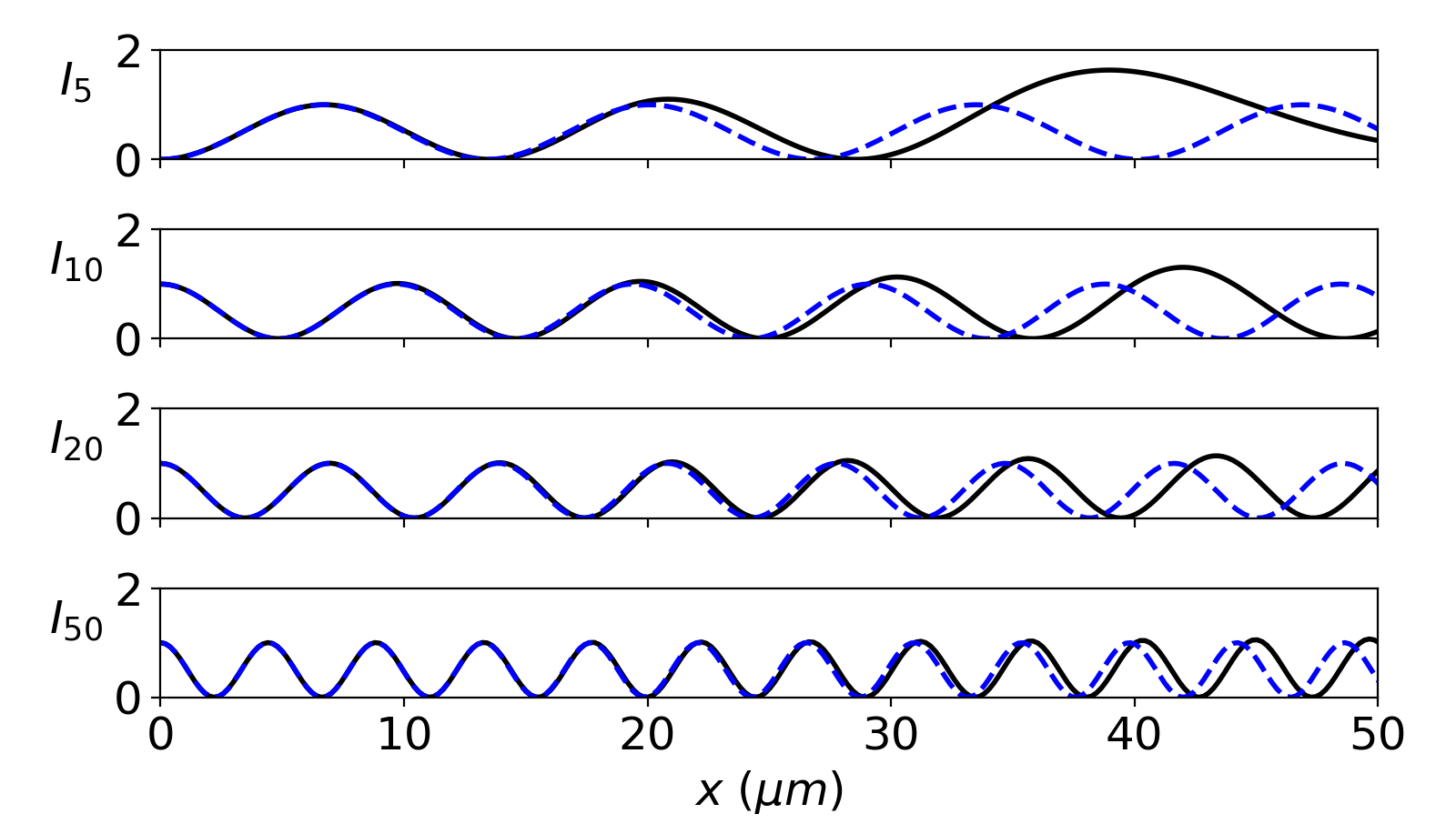}
	\caption{\textcolor{black}{Figure 3.3 from~\cite{moulanier2024modelisation}. 
	Evolution of $I_m$, the intensity of a 1D HG mode defined along the $x$ direction, for $w_{0}=20$ \si{\micro\metre} and $m\in\left[5,\,10,\,20,\,50\right]$. $I_m$ is normalized by its first local maximum along $x \geq 0$.
    For each plot, the black curve represents the calculation of $I_m$ given by the analytical expression of an HG mode, and the dashed blue curve is the expected asymptotic behaviour of $I_m$ given by Eq. (\ref{eq:HG_cos_scale})~\cite{dominici2007asymptotic}.}}
	\label{fig:HG_cosine}
\end{figure}

\textcolor{black}{
As shown in Fig.\,\ref{fig:HG_cosine}, for $m \gg 1$ and $x \leq w_{0}/\sqrt{2}$, higher orders modes can be approximated with a periodic function~\cite{dominici2007asymptotic, moulanier2024modelisation}:
\begin{equation}\label{eq:HG_cos_scale}
    I_m \propto \cos^2 \left(X\sqrt{2m+1} - \dfrac{m\pi}{2}\right),
\end{equation}
where $X = \sqrt{2}x/w_0$. Based on Eq.\,(\ref{eq:HG_cos_scale}), the larger mode order oscillates at the highest frequency among $\set{I_m(x,z_f)}$. Consequently, a lower bound for $\delta r$ is the distance between one local maximum and the next local minimum of $I_{N}(x, z_f)$:
\begin{equation}\label{eq:HG_resolution}
    \begin{aligned}
        \delta r & \gtrsim \dfrac{w_0}{\sqrt{2}}|X_{min,k} - X_{max,k}|\\
                 & \gtrsim \dfrac{w_0}{\sqrt{N+\dfrac{1}{2}}}\dfrac{\pi}{4},
    \end{aligned}
\end{equation}
where $X_{min,k}$ and $X_{max,k}$ are respectively the $k$-th local minimum and maximum of Eq.\,(\ref{eq:HG_cos_scale}) with $m=N$.
}

\textcolor{black}{
To summarize, since $\delta r \propto w_0 / \sqrt{N}$ and $r_{E} \propto w_0 \sqrt{N}$, setting \{$w_0$, $N$\} based on the size of the signal domain and the minimum size of details to reconstruct will control both the extent of the HG mode basis and its resolution.
}

\textcolor{black}{
Since $|\mathrm{HG}_{m,n}|^2$ is fully separable in $x$ and $y$, $\delta r$ and $r_{E}$ can also be extended to a 2D HG basis with $w_{0,x} \neq w_{0,y}$ and $N_m \neq N_n$ by reasoning with the parameter pairs \{$N_m$, $w_{0,x}$\}, \{$N_n$, $w_{0,y}$\}. 
}

\textcolor{black}{
For the LG basis, we derive parametric dependence of $\delta r$ and $r_E$ close to the HG case. For $l=0$, each cylindrically symmetric modes of the LG basis can be described as a sum of 2D HG modes orders up to $2N_p$ in $x$ and in $y$~\cite{kimel2002relations}. Hence, $r_{E}$ is also expected to scale with $w_0\sqrt{N}$ for LG modes.
For $N \rightarrow +\infty$, Laguerre polynomials can be approximated with Bessel functions~\cite{temme1990asymptotic, deano2013strong}:
\begin{equation}\label{eq:delta_r_LG}
     \lim_{N \rightarrow \infty}l^0_N\left(\frac{z^2}{4N}\right) = J_0(z),
\end{equation}
where $l_p^0(z)$ is the generalized Laguerre polynomial of azimuthal order $l=0$ and radial order $p$, and $J_0$ the Bessel function of first kind, $J_l$, with $l=0$~\cite{abramowitz1966handbook}. In the explicit expression of the LG basis from Eq.\,(16) of~\cite{massimo2025laser}, at focus, $\mathrm{LG}_{0,p}$ is a function of $l_p^0(2r^2/w_0^2)$. Replacing $z$ with $2\sqrt{2N}r/w_0$ in Eq.\,(\ref{eq:delta_r_LG}), with $N$ sufficiently large, the positions of the first zeros of the cylindrically symmetric modes of the LG basis are expected to scale with $w_0/\sqrt{N}$, similarly to $\delta r$ in the HG basis.
}

\textcolor{black}{
In the use case of the paper, $\delta r \sim 5$ \si{\micro\metre} and $r_{E} \sim 165$ \si{\micro\metre}.}

\begin{figure}[h]
	\centering
	  \includegraphics[width=0.9\linewidth]{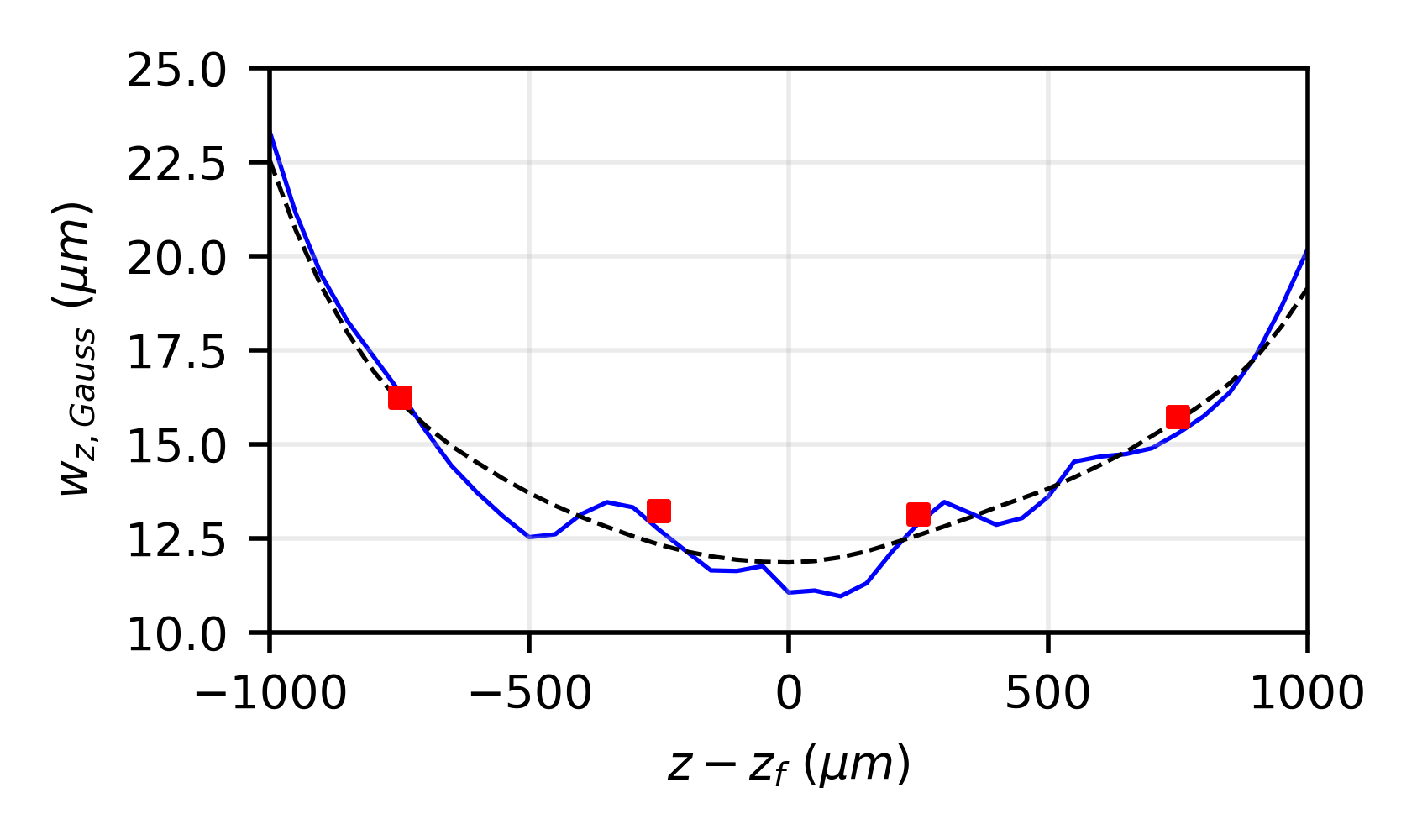}
	\caption{\textcolor{black}{Evolution of $w_{z, Gauss}$, the laser waist determined by a Gaussian fit of the measured laser fluence distribution at each transverse plane, as a function of $z-z_f$, the laser relative position along the propagation axis.
    Red squares correspond to values determined on laser fluence images, dashed black curve to values predicted by the GSA-MD with an HG basis using $N_{m,n}=N=30$, $w_{0,x,y}=w_{0}=30$ \si{\micro\metre}, blue curve to values predicted by the GSA-MD with an HG basis using $N=30$ and $w_{0}=13.2$ \si{\micro\metre}.}}
	\label{fig:waists}
\end{figure}

\textcolor{black}{
The numerical waist of the HG/LG modes, $w_{0}$, is different from the physical waist that can be estimated by transverse diagnostics such as in Fig.\,\ref{fig:Waist_Lund}: the former is a user-defined parameter of the GSA-MD reconstruction that controls the characteristic lengths of the modes in Eq.\,(\ref{eq:Efield_spatial}) (without affecting the total energy defined in Eq.\,(\ref{eq:energy})), while the latter is the typical radius over which the laser intensity at focus drops to $1/e^2$ of its peak value.
This distinction is important for a multi-modal reconstruction such as the GSA-MD, since smoother modes ($w_{0}$ greater than the physical waist) can lead to a less noise-sensitive reconstruction due to the increase in $\delta r$~\cite{thevenet2025lasy}.
}

\textcolor{black}{
Furthermore, the modes radial extent scales with $w_{0}$, so if its value is too low, higher order modes will capture a significant share of the laser energy. In that case, these higher orders of the HG/LG decomposition can dominate the diffraction of the reconstructed laser, which has a characteristic length $z_R = w^2_{0}k/2$ across all modes. 
}

\textcolor{black}{
In the example presented in Fig.\,\ref{fig:waists}, the physical waist fitted by a Gaussian distribution at $z^*-z_f=-250$ \si{\micro\metre} was $w_{z^*, Gauss}=13.2$ \si{\micro\metre}. Using $w_0=w_{z^*, Gauss}$ for the GSA-MD in HG basis leads to a great agreement with the experiment at the positions of the experimental fluence measurements (red squares).
However, this comes at the cost of overfitting the data and introducing fluctuations in the intermediate planes. By choosing a larger numerical waist, $w_{0}=30$ \si{\micro\metre}, the quality of the reconstruction is lesser in measurement planes, but the curve of $w_{z,Gauss}$ in intermediate planes is much smoother and resembles the expected behaviour of a Gaussian-like laser, thus preventing overfitting.
}

\textcolor{black}{\subsection{Modelling of the transverse distribution in PIC simulations}\label{sec:appendix_transverse_distribution}}

\textcolor{black}{
This appendix details the modelling of the ACE transverse distribution in PIC simulation, and the benefits of using a realistic distribution obtained via GSA-MD over a standard Gaussian fit in a PIC simulation.
}

\textcolor{black}{
To perform the ``Basis Swap" referred in Fig.\,\ref{fig:Framework}, conversion operations between HG and LG modes have been demonstrated in~\cite{kimel2002relations}.
While lossless, this option restricts the choice of LG modes that can be generated from a truncated HG basis with $N_m+1$ and $N_n+1$ modes. 
A more straightforward procedure was presently adopted: the distribution of the complex GSA-MD electric field reconstructed in its focal plane is projected over the basis of normalized LG modes. The electric field $\tilde{E}_{\perp}$ satisfies Eq.\,(\ref{eq:Efield_spatial}) for the HG basis, and is assumed to have a truncated expression in the LG basis. Consequently, $\tilde{E}_{\perp, \mathrm{LG}}$, the transverse electric field in the LG mode basis, $\tilde{E}_{\perp, \mathrm{LG}}$, is obtained by a sum of projections at focus:
\begin{equation}\label{eq:projectionHG}
    \begin{aligned}
        \tilde{E}_{\perp, \mathrm{LG}}(r,z_f,\theta) &= \sum_{l,p}^{N_l,N_p} \mathrm{LG}_{l,p}(r,z_f,\theta)\mathrm{Proj} [\tilde{E}_{\perp},\,\mathrm{LG}_{l,p}]\\
        &= \sum_{l,p}^{N_l,N_p} \tilde{C}_{l,p}\mathrm{LG}_{l,p}(r,z_f,\theta),
    \end{aligned}
\end{equation}
where $\mathrm{LG}_{l,p}$ is the normalized LG mode of azimuthal order $l$ and radial order $p$ so that $\|\mathrm{LG}_{l,p}\|^2=1$, and $\mathrm{Proj}$ is the projection operator over a normalized mode of the LG basis:
\begin{equation}
    \mathrm{Proj}[f,\mathrm{LG}_{l,p}](z)=\iint f(r,z,\theta) \mathrm{LG}_{l,p}^*(r,z,\theta)\mathop{r} \mathop{dr} \mathop{d\theta}.
\end{equation}
}

\textcolor{black}{
Compared to the mode conversions in~\cite{kimel2002relations}, this method allows the user to choose the number of azimuthal and radial modes in the output LG basis more freely. The cost of this is that a residual error arises from the projection, which can be non-negligible if the features of the truncated LG basis, e.g. $\delta r$ and $r_{E}$, are too far from those of the HG basis. Hence, monitoring the quality of the reconstructed distribution after the conversion to the LG mode basis helps the user to set $N_l$, $N_p$ and $w_0$ accordingly.
}

\begin{figure}[h]
	\centering
	  \includegraphics[width=0.9\linewidth]{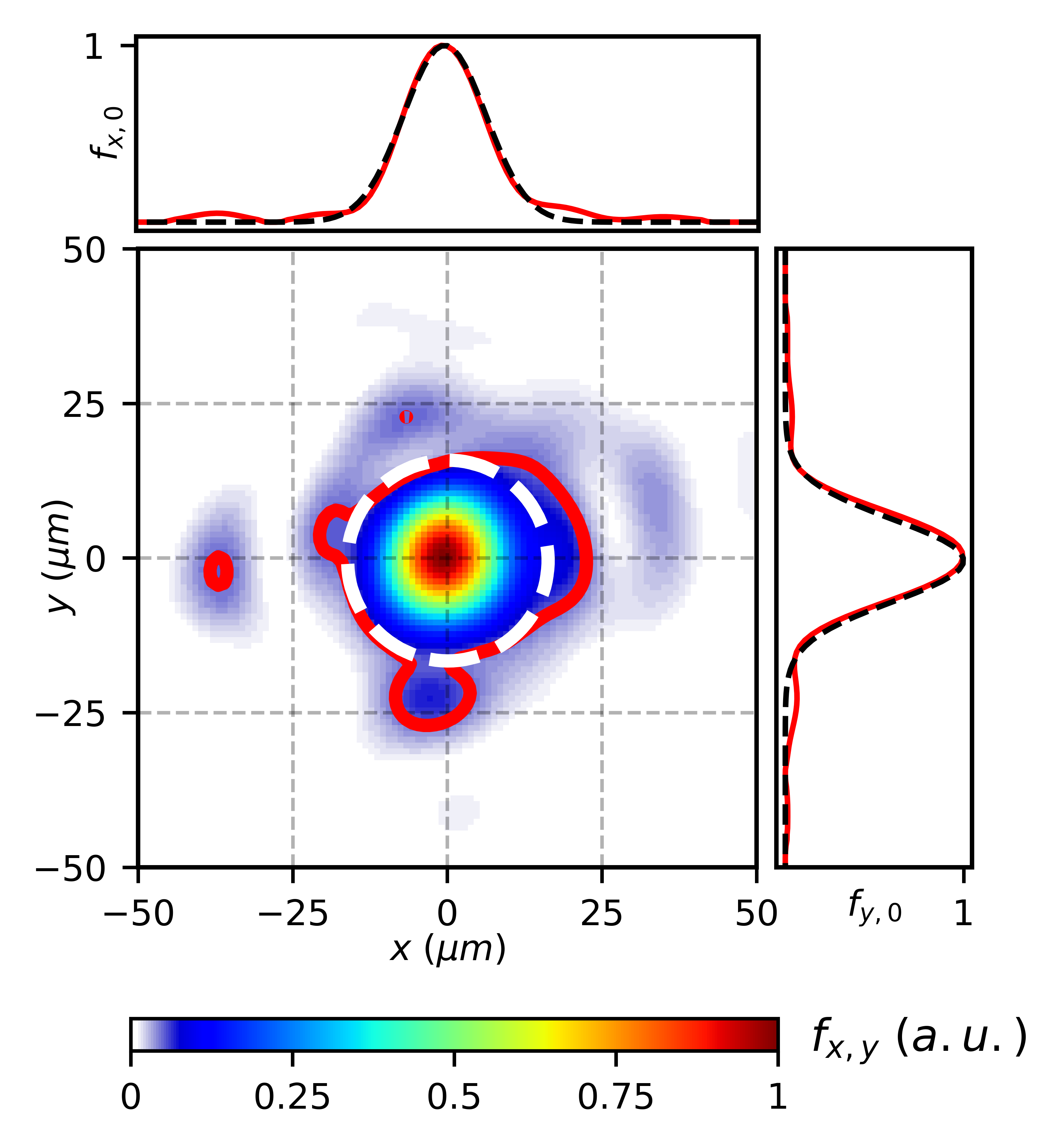}
	\caption{\textcolor{black}{Distribution of the normalized laser fluence, $f_{x,y}$, measured in vacuum in the LLC experiment at $z^*-z_f=-250$ \si{\micro\metre}.
    The dashed white circle includes the fraction of the circularly symmetric Gaussian fit ($1/e^2$ waist $w_{Gauss}=13.2$ \si{\micro\metre}) above $5$\,\si{\percent} of its maximum. The red dashed line is the same limit for the measured fluence distribution.
    $f_{x,0}$ and $f_{0,y}$, the fluence line profiles at $y=0$ and $x=0$, are plotted for the Gaussian fit and experimental distribution as a black dashed line and red solid line, respectively.}}
	\label{fig:Gauss_fit}
\end{figure}

\textcolor{black}{
Within the GSA-MD, the reconstructed distribution quality is evaluated by the integrated residuals~\cite{moulanier2023fast, massimo2025laser}:
\begin{equation}\label{eq:error_chi_fluence}
\chi^2_{\operatorname{GSA-MD}}=\sum_{k=0}^{N_z}\dfrac{\sqrt{\sum_{i,\thinspace j}\left(f_{norm} - |\tilde{E}_{\perp, norm}|^2\right)^2(x_i, y_j, z_k)}}{(N_z + 1)\sum_{i,\thinspace j} f_{norm}(x_i, y_j, z_k)},   
\end{equation}
where the subscript ``$norm$" indicates that both transverse distributions area integrals have been normalized to the same value, $k\in[0,1,...,N_z]$ is the index of the discrete set of measurement planes, $i$ the transverse index of the $k$-th grid in the $x$ direction and $j$ in the $y$ direction.
For the projection error, we use the average least absolute deviation~\cite{powell1984least}, which is a metric similar to Eq.\,(\ref{eq:error_chi_fluence}) but more sensible in low signal areas.
The error is calculated between $F_{\mathrm{LG}}$, the fluence of the LG projection sum, and $F_{\mathrm{HG}}$, the fluence of the original GSA-MD transverse electric field in the HG basis. For two discrete distributions $G$ and $H$ defined over the same domain, we define $S_\beta$, a filtered version of the averaged least absolute deviation:
\begin{equation}\label{eq:error_chi_proj}
    \begin{aligned}
        &S_\beta(z) = \dfrac{1}{\sum_{i,\thinspace j} \mathbbm{1}_{\operatorname{res}}(x_i, y_j, z)}\sum_{i,\thinspace j} \operatorname{res}(x_i, y_j, z)\\
        &\operatorname{res} =
        \set[\bigg]{\dfrac{| G - H|}{G} \given G\geq \beta\times\max_z\left(G\right)},
    \end{aligned}    
\end{equation}
where $\mathbbm{1}_{\operatorname{res}}$ is the indicator function of the residuals that equals 1 on points for which a residual is defined and 0 otherwise, $\beta\in [0,1)$ and $\max_z\left(G\right)$ the global maximum of $G$ at the position $z$.
As shown in Fig.\,\ref{fig:Gauss_fit} by the red line, for the LLC use case, setting $\beta=0.05$ ensures that the calculation is done above noise-level, while still factoring the asymmetries of the transverse distribution.
We define the projection error at focus $S_{\operatorname{Proj}}$ as Eq.\,(\ref{eq:error_chi_proj}) calculated with $G=F_{\mathrm{HG}}$, $H=F_{\mathrm{LG}}$, $\beta=0.05$ and $z=z_f$, the plane in which the projection of Eq.\,(\ref{eq:projectionHG}) was performed.
For the transverse distribution used in this paper, i.e. $N_m=N_n=30$, $w_0=30$ \si{\micro\metre} for the HG basis and $N_p=30$, $N_l=3$, $w_0=30$ \si{\micro\metre} in the LG basis, $S_{\operatorname{Proj}}=1.8$\,\si{\percent}. As a reference, replacing the LG distribution in Eq.\,(\ref{eq:error_chi_proj}) with the fluence of a fitted circularly symmetric Gaussian distribution with waist $w_{0,Gauss}=12.1$ \si{\micro\metre} yields $S_{\operatorname{Proj}}=23$\,\si{\percent}.
$S_\beta(z)$ was also calculated over $z \in \left[-750,\,+750\right]$ \si{\micro\metre}, which is the span covered by the 4 measurement planes used in the GSA-MD, following Eq.\,(\ref{eq:error_chi_proj}) and replacing $z_f$ with $z$. Using increments of $100$ \si{\micro\metre} in $z$, the residual error averaged over this interval reaches $2.2$\,\si{\percent} for the LG projection and $42$\,\si{\percent} for the Gaussian fit. This confirms that the LG field from Eq\,(\ref{eq:projectionHG}) preserves the features of the original HG decomposition close to focus.
}

\textcolor{black}{
The importance of $\alpha$, the energy ratio obtained in Eq.\,(\ref{eq:E_ratio}) of Sec.\,\ref{sec:Implementation}, is illustrated by Fig.\,\ref{fig:Gauss_fit} for a standard case where the modelled distribution is not a perfect reconstruction. This figure shows a comparison between the distributions of the measured laser transverse distribution at $z^*-z_{f}=-250$ \si{\micro\metre} (second data point in Fig.\,\ref{fig:Waist_Lund}), and a fitted circularly symmetric Gaussian distribution with $1/e^2$ waist $w_G=13.2$ \si{\micro\metre}. The area delimited by a red (resp. white) line represents the fraction of the experimental (Gaussian fit) distribution which amplitude is above $5$\,\si{\percent} of its maximum. The difference between the two indicates that $\alpha < 100$\,\si{\percent}.
}

\textcolor{black}{
In this figure, $\alpha$ was calculated in Eq.\,(\ref{eq:a0_ratio}) by substituting $|\tilde{e}_\perp|$ with the circularly symmetric Gaussian distribution of waist $w_G$ and peak value of $1$, yielding $\alpha=79$ \si{\percent}. Using more refined methods of transverse reconstruction such as the LG decomposition usually increase the value of $\alpha$ towards $100$\,\si{\percent} as demonstrated in Sec.\,\ref{sec:Implementation} where $\alpha=85$\,\si{\percent} was evaluated. In comparison to standard fitting methods using one measurement plane, the GSA-MD distributions also tend to diminish the pixel-to-pixel error with respect to the experiment. To demonstrate the second point, we calculated $S_{\beta}$ at $z=z^*$ with $G=F(x,y,z^*)$, the measured laser fluence and $H=F_{\mathrm{LG}}(x,y,z^*)$. We compared it to the same metric using $H=F_{Gauss}$, the laser fluence of the Gaussian fit from Fig.\,\ref{fig:Gauss_fit}. With $\beta=0.05$, $S_{\beta}(z^*)=9.9$ \si{\percent} for the LG decomposition while $S_{\beta}(z^*)=17.7$ \si{\percent} for the circularly Gaussian fit.}

\begin{figure}[h]
	\centering
	  \includegraphics[width=0.9\linewidth]{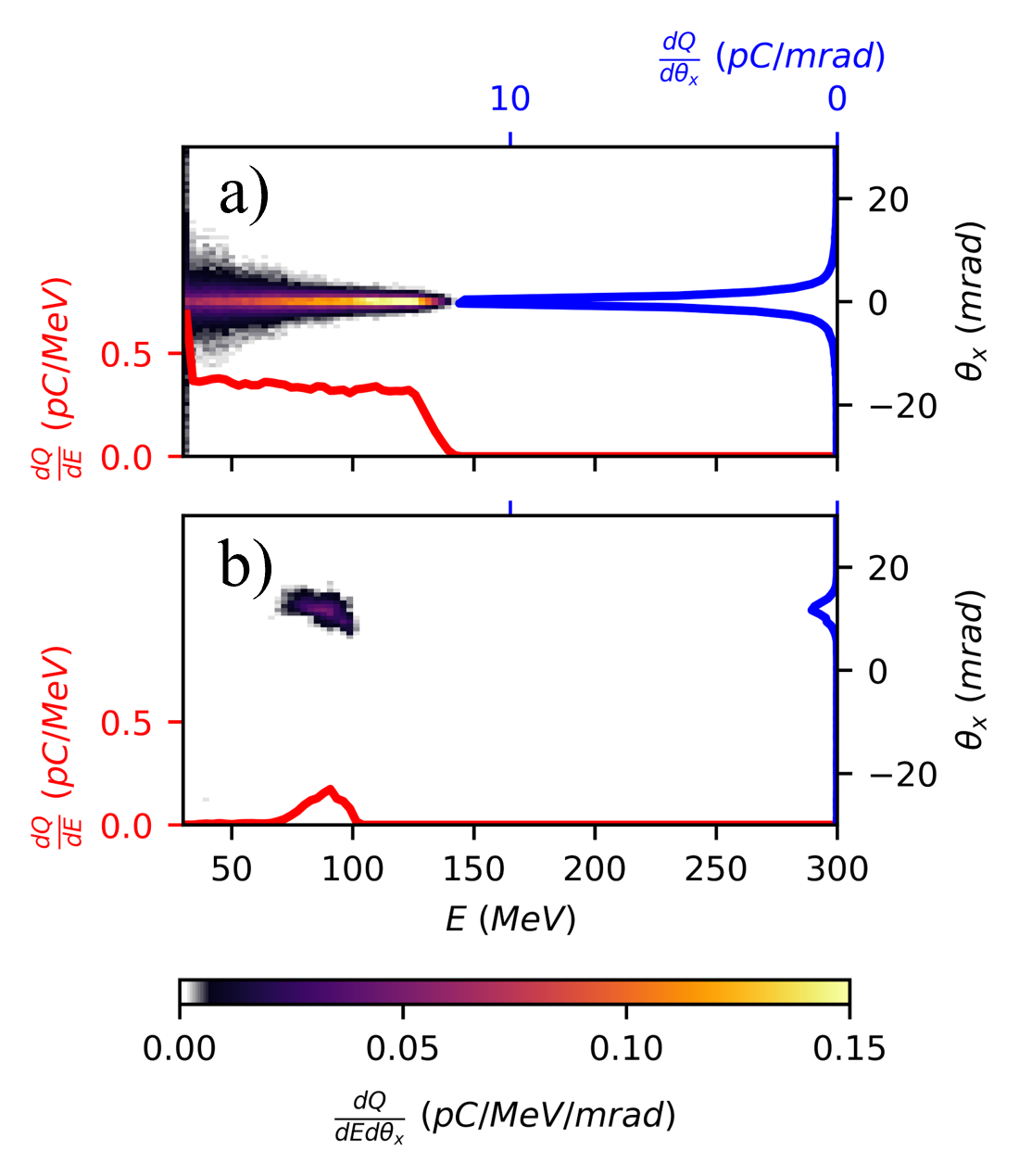}
	\caption{\textcolor{black}{Electron spectrum at the plasma exit in two PIC simulations using the laser temporal profile of the TCS and different transverse distributions:
	a) circularly symmetric Gaussian fit from Fig.\,\ref{fig:Gauss_fit} (dashed circle in white) with $w_{0,Gauss}=13.2$\,\si{\micro\metre} - b) LG decomposition used for the TCS with $N_l=3$, $N_p=30$ and $w_0=30$\,\si{\micro\metre}.}}
	\label{fig:Transverse_effects}
\end{figure}

\textcolor{black}{
In the context of transverse modelling using GSA-MD over a perfect Gaussian approximation, correlations between the pointing of the laser, its fluctuations, and the energy gain and electron bunch pointing have been previously established~\cite{moulanier2023fast, moulanier2025coupled}.
Fig.\,\ref{fig:Transverse_effects} illustrates the same effects of the transverse distribution of ACE lasers. In this figure, the output electron bunch is represented in the $E-\theta_x$ plane at the plasma exit for two simulations sharing the same chirped temporal profile as the TCS, but two different transverse distributions: the fitted Gaussian distribution from Fig.\,\ref{fig:Gauss_fit} with $w_{0,Gauss}=13.2$ \si{\micro\metre} [Fig.\,\ref{fig:Transverse_effects}.(a)], and the LG decomposition with $N_l=3$, $N_p=30$, $w_0=30$ \si{\micro\metre} that was used for the TGS, TCS and TAS [Fig.\,\ref{fig:Transverse_effects}.(b)]. Even after factoring the value of $\alpha$ to calibrate the Gaussian fit energy, the bunch charge, energy gain and energy spread in Fig.\,\ref{fig:Transverse_effects}.(a) are over-estimated with respect to the experimental profile in Fig.\,\ref{fig:chirp_dQdE} (black dashed line). Furthermore, the off-axis pointing of the electron bunch, which was $8$ \si{\milli\radian} in the experiment~\cite{Backhouse2025} and $12$ \si{\milli\radian} in Fig.\,\ref{fig:Transverse_effects}.(b) for the TCS, cannot be reproduced with an axisymmetric transverse distribution. This limitation stems from the Gaussian laser fluence remaining centered on its propagation axis in the plasma~\cite{moulanier2023modeling, moulanier2025coupled}.}

\textcolor{black}{\subsection{Simulation parameters}\label{sec:appendix_sims}}

\textcolor{black}{
This appendix details the reasoning behind the choice of several numerical parameters of simulations shown in Table\,\ref{tab:Lund_sims_parameters} of Sec.\,\ref{sec:sim_inputs}.}

\textcolor{black}{
In the quasi-3D cylindrical geometry, the choice of the number of azimuthal modes in the Pseudo-Spectral Analytical Time-Domain (PSATD) solver of Maxwell equations~\cite{vay2013domain, godfrey2014numerical} depends on two competing factors. First, the quality of the reconstruction of an asymmetric pulse typically increases with $N_l$. Secondly, increasing $N_\theta$, the maximum azimuthal order for the harmonic decomposition operated in the PSATD, demands a co-linear increase in the number of macro-particles per $2\pi$ angle to accurately sample the field distributions~\cite{lifschitz2009particle}.
}

\textcolor{black}{
Since multiple steps of a PIC cycle are performed on the macro-particles~\cite{BirdsallLangdon}, increasing their number rises the simulation running time. To fully model a linearly polarized laser over an LG basis with a fixed maximum azimuthal order $N_l$, a maximum harmonic order $N_\theta=N_l+1$ is required for the PSATD solver. This value equates the maximum azimuthal order from the LG basis, plus one from the multiplication by terms $[\cos(\theta), \sin(\theta)]$ rising from the projection to unitary cylindrical vectors $\overrightarrow{u}_r$, $\overrightarrow{u}_\theta$~\cite{lifschitz2009particle, Lehe2016}.
}

\textcolor{black}{
While the experimental laser energy is $E_{laser}=0.87$ \si{\joule}, the modelled laser energy was set to $E_{laser,sim}=0.74$ \si{\joule} in order to account for the energy ratio $\alpha=85$ \si{\percent} (Eq.\,(\ref{eq:a0_ratio}) of Sec.\,\ref{sec:Implementation}).
}

\textcolor{black}{
The longitudinal plasma density profile was modelled based on OpenFOAM simulations of the ELISA gas cell~\cite{audet2018gas} used for the LLC experiment. The gas composition was modelled with $95$ \si{\percent} of $\mathrm{H_2}$ doped by $5$ \si{\percent} of $\mathrm{N_2}$.
All PIC simulations were performed within the Lorentz boosted frame~\cite{vay2007noninvariance, yu2016enabling}.
}

\textcolor{black}{
Compared to an idealized bi-Gaussian approximation of the laser spatio-temporal distribution, PIC simulations of an ACE laser generally require higher resolutions. However, the simulation cost can be leveraged while maintaining the accurate shape of the modelled laser. In regimes where the chirping of frequencies within the temporal profile are negligible, an envelope model can be adopted to represent $|\tilde{T}(t)|$~\cite{massimo2020numerical}. Furthermore, resolution parameters inherent to PIC simulations such as the number of macro-particles per angle $P_\theta$ and the grid radial step size $\Delta r$ can both be downscaled by decreasing the number of modes --- $N_l$ and $N_p$ for LG or $N_m$ and $N_n$ for HG --- and by simultaneously increasing $w_0$, the numerical waist of the GSA-MD, thus smoothing the radial gradient of the modal decomposition.
}

\vspace{10mm}
\subsection{Fitting the bunch maximum energy with $\phi_2$}\label{sec:appendix_fit_E}

This appendix describes the equations used to fit the data points plotted in Fig.\,\ref{fig:chirp_scan}.(a) for the bunch maximum energy $E_{max}$.
\textcolor{black}{
The linear scaling between the electron energy gain and the laser electric field amplitude is observed in the bubble regime which arises for $a_0 \geq 2$~\cite{lu2007generating}. It is characterized by a spherical plasma cavity inside which the longitudinal electric field becomes linear~\cite{kostyukov2004phenomenological, lu2007generating}. In the simulations of this paper, $a_0$ reached nonlinear values between $1.5$ and $1.9$, which is close to the expected threshold for the bubble regime.}
We modelled the energy gain with a linear relation between $E_{max}$ and $a_{0}$, as previously demonstrated in the 3D nonlinear bubble regime~\cite{lu2007generating, couperus2017demonstration}. In our simplified model, where $\phi_2$ diminishes the chirped amplitude to $a_{0,c}=a_0\chi^{-1/4}$ according to Eq.\,(\ref{eq:a_chirp}), the bunch maximum energy is negatively correlated with $\phi_2$.
The points of Fig.\,\ref{fig:chirp_scan}.(a) were fitted with the following curve:
\begin{equation}\label{eq:E_fit}
    E_{max} = A_1 \times a_{0,c}(\phi_2)+C_1,
\end{equation}
where $C_1 \in \mathbb{R}$ and $A_1>0$ are the fit constants to be determined by the nonlinear least squares method~\cite{lawson1995solving}. The optimal fit parameters using Eq.\,(\ref{eq:E_fit}) led to a coefficient of determination of $R^2=0.99$ with $A_1=223.7$ \si{\mega\electronvolt} and $C_1=-262.8$ \si{\mega\electronvolt}, which indicates a near-zero sum of squared residuals~\cite{steel1960principles}.
In the fit of Eq.\,(\ref{eq:E_fit}), $\tau_l$ and $a_0$ were replaced by the parameters used for simulations, i.e. unchirped laser duration $\tau_l=33.3$ \si{\femto\second} and the maximum recorded laser amplitude for the TGS, $a_0=1.9$.

\subsection{Fitting the bunch charge and energy spread with $\phi_2$}\label{sec:appendix_fit_sigmaE}

This appendix describes the equations used to fit the data points plotted in Fig.\,\ref{fig:chirp_scan}.(b) for the charge $Q$, and Fig.\,\ref{fig:chirp_scan}.(c) for the energy spread $\sigma_E$.
The bunch injected charge, $Q$, grows with the volume in which the laser envelope is sufficiently high for the ionization of $N^{5+}$. Since the input chirp coefficients only affect $\tilde{T}(t)$ [Eq.\,(\ref{eq:Efield_time})] and normalized amplitude $a_{0,c}$ [Eq.\,(\ref{eq:A_inj})], the variation of the volume of ionization with $\phi_2\neq0$ is a function of the integral of $a(t)=a_{0,c}\times |\tilde{T}(t)|$ inside the interval where $a(t)$ exceeds the threshold of $N^{5+}$ ionization. Theoretically, the threshold amplitude sits at $a(t)\sim1.5$~\cite{chen2012theory}.
In the following, the temporal integral of $a(t)$ that exceeds the ionization threshold of $N^{5+}$ will be written $\mathcal{A}_{i}$. 
The points of Fig.\,\ref{fig:chirp_scan}.(b) were fitted with a simple linear correlation between $Q$ and $\mathcal{A}_{i}(\phi_2)$:
\begin{equation}\label{eq:sigmaE_fit}
    Q = A_2 \times \mathcal{A}_{i}(\phi_2)+C_2,
\end{equation}
where $C_2\in \mathbb{R}$ and $A_2>0$ are the fit constants.

To measure the role of $\mathcal{A}_{i}$ in the amount of injected charge, the integral $\mathcal{A}_{i}(\phi_2)$ was evaluated for $z-z_{cell}=300$ \si{\micro\metre}. This plane corresponds to the middle of the injection phase that spans from $z-z_{cell}=100$ to $500$ \si{\micro\metre} and is also the focal plane position of the laser within the plasma in Fig.\,\ref{fig:Lund_sims_comparison}.(a).
In Eq.\,(\ref{eq:sigmaE_fit}), $\mathcal{A}_{i}$ was calculated by estimating the integral of $a(t)$ for $t\in[t_{i,0},t_{i,1}]$ where $t_{i,0,1}$ are the time coordinates in between which $a(t\in[t_{i,0},t_{i,1}]) \gtrsim 1.5$~\cite{chen2012theory}. Assuming that the constant contributions of $\phi_3$ and $\phi_4$ are taken into account within $A_2$ in Eq.\,(\ref{eq:sigmaE_fit}), the chirped profile $\mathcal{A}_{i}$ is calculated as a function of $\phi_2$ only. In this approximation, the envelope of $\real\{\tilde{T}(t)\}$ is a symmetric Gaussian profile with $1/e^2$ duration $\tau_{l,c}$ parameterized by $\phi_2$ [Eq.\,(\ref{eq:tau_chirp})]. In this case, $t_{i,0}=-t_{i,1}=\pm t_{i}$, with $t_{i}$ retrieved by:
\begin{equation}\label{eq:t_inj}
    \begin{aligned}
        &a(t_{i})\simeq1.5\\ 
        &a_{0}\,\chi^{-1/4}\,\exp\left[-t^2_{i}/(\eta\tau_{l,c})^2\right]\simeq1.5\\
        &\exp\left[-t^2_{i}/(\eta\tau_{l,c})^2\right]\simeq \chi^{1/4} \frac{1.5}{a_{0}}\\
         &\frac{1}{\eta^2}\left(\frac{t_{i}}{\tau_{l,c}}\right)^2 \simeq -\frac{\ln\left(\chi\right)}{4}+\ln(a_{0}) - \ln(1.5)\\
         &t_{i}\simeq \pm \eta\tau_{l,c}\sqrt{-\frac{\ln\left(\chi\right)}{4}+\ln(a_{0}) - \ln(1.5)},
    \end{aligned}
\end{equation}
where $\eta$ is a constant which accounts for the compression of the laser duration within the plasma compared to the duration at the start of simulation. In the middle of the injection ($z-z_{cell}=300$ \si{\micro\metre}), $\eta$ was evaluated to be $0.81$ in the TCS [Fig.\,\ref{fig:Lund_sims_comparison}.(c), blue points] as well as in the TGS (green points).
Since values of $\phi_2$ ranging from $0$ \si{\femto\second\squared} (value of the TGS) to $501$ \si{\femto\second\squared} (value of the TCS) are expected to produce similar compression factors, the value of $\eta$ was also set to $\eta=0.81$ for the fit.
Setting the boundaries found in Eq.\,(\ref{eq:t_inj}) yields the approximated ionization integral $\mathcal{A}_{i}(\phi_2)$:
\begin{equation}\label{eq:A_inj}
    \begin{aligned}
        \mathcal{A}_{i}(\phi_2) 
        &= a_{0}\,\chi^{-1/4}\int_{-t_{i}}^{+t_{i}} e^{-t^2/(\eta\tau_{l,c})^2} \mathop{dt}\\
        &= a_{0}\,\eta\tau_l\,\sqrt{\pi}\,\chi^{1/4}\,\mathrm{erf}\left(\frac{t_{i}}{\eta\tau_{l,c}}\right)\\
        &\simeq a_{0}\,\eta\tau_l\,\sqrt{\pi}\,\chi^{1/4}\\
        &\hspace{15pt}\times\mathrm{erf}\left(\sqrt{-\frac{\ln\left(\chi\right)}{4}+\ln(a_{0}) - \ln(1.5)}\right)\\
    \end{aligned}
\end{equation}
where $\mathrm{erf(x)}$ is the Gauss error function~\cite{andrews1998special}.
In simulations, the propagation span over which injection occurs coincides with the interval where $a_0$ reaches its maximum [see Figs.\,\ref{fig:Lund_sims_comparison}.(a) and (d)]. Therefore, the value of $a_0$ that was used for the fit of Eq.\,(\ref{eq:A_inj}) is the maximum recorded for the TGS, $a_0 \simeq 1.9$. The $1/e^2$ duration was set to that of the unchirped laser with $\tau_l=33.3$ \si{\femto\second}.
The fit yields $R^2=0.98$ with $A_2=2.39$ \si{\pico\coulomb\per\second} and $C_2=-76.2$ \si{\pico\coulomb}, which confirms a good agreement with the simulated data points. The bunch energy spread, $\sigma_E$, is strongly correlated with the volume of injected charge. To fit the points of Fig.\,\ref{fig:chirp_scan}.(c), we replaced $Q$ with $\sigma_E$ in Eq.\,(\ref{eq:sigmaE_fit}). With $A_2=1.63$ \si{\mega\electronvolt\per\second} and $C_2=-42.3$ \si{\mega\electronvolt}, $R^2=0.95$.

\vspace{0.25cm}

\bibliography{bib}

@article{kim2017stable,
  title={Stable multi-GeV electron accelerator driven by waveform-controlled PW laser pulses},
  author={Kim, Hyung Taek and Pathak, VB and Hong Pae, Ki and Lifschitz, Agustin and Sylla, Fran{\c{c}}ois and Shin, Jung Hun and Hojbota, C and Lee, Seong Ku and Sung, Jae Hee and Lee, Hwang Woon and others},
  journal={Scientific reports},
  volume={7},
  number={1},
  pages={10203},
  year={2017},
  publisher={Nature Publishing Group UK London}
}

@book{Siegman86,
  author = {Siegman, Anthony E.},
  publisher = {University Science Books},
  title = {Lasers},
  year = 1986
}

@article{moulanier2023fast,
  title={Fast laser field reconstruction method based on a Gerchberg--Saxton algorithm with mode decomposition},
  author={Moulanier, Ioaquin and Dickson, Lewis Thomas and Massimo, Francesco and Maynard, G and Cros, B},
  journal={JOSA B},
  volume={40},
  number={9},
  pages={2450--2461},
  year={2023},
  publisher={Optica Publishing Group}
}

@article{kimel2002relations,
  title={Relations between hermite and laguerre gaussian modes},
  author={Kimel, Isidoro and Elias, Luis R},
  journal={IEEE Journal of quantum electronics},
  volume={29},
  number={9},
  pages={2562--2567},
  year={2002},
  publisher={IEEE}
}

@phdthesis{Dickson2023,
  TITLE = {{Laser wakefield acceleration of electrons}},
  AUTHOR = {Dickson, Lewis},
  URL = {https://theses.hal.science/tel-04606600},
  NUMBER = {2023UPASP037},
  SCHOOL = {{Paris-Saclay University}},
  YEAR = {2023},
}

@article{Lehe2016,
	author = {R{\'e}mi Lehe and Manuel Kirchen and Igor A. Andriyash and Brendan B. Godfrey and Jean-Luc Vay},
	doi = {https://doi.org/10.1016/j.cpc.2016.02.007},
	issn = {0010-4655},
	journal = {Computer Physics Communications},
	pages = {66-82},
	title = {A spectral, quasi-cylindrical and dispersion-free Particle-In-Cell algorithm},
	url = {https://www.sciencedirect.com/science/article/pii/S0010465516300224},
	volume = {203},
	year = {2016}}

@article{Sroor2021,
	author = {Hend Sroor and Chane Moodley and Valeria Rodr\'{i}guez-Fajardo and Qiwen Zhan and Andrew Forbes},
	doi = {10.1364/JOSAA.432431},
	journal = {J. Opt. Soc. Am. A},
	month = {Oct},
	number = {10},
	pages = {1443--1449},
	publisher = {Optica Publishing Group},
	title = {Modal description of paraxial structured light propagation: tutorial},
	url = {https://opg.optica.org/josaa/abstract.cfm?URI=josaa-38-10-1443},
	volume = {38},
	year = {2021}}

@article{Jones1998,
	author = {Jones, Donald R. and Schonlau, Matthias and Welch, William J.},
	date = {1998/12/01},
	doi = {10.1023/A:1008306431147},
	id = {Jones1998},
	isbn = {1573-2916},
	journal = {Journal of Global Optimization},
	number = {4},
	pages = {455--492},
	title = {Efficient Global Optimization of Expensive Black-Box Functions},
	url = {https://doi.org/10.1023/A:1008306431147},
	volume = {13},
	year = {1998}}

@misc{Frazier2018,
      title={A Tutorial on Bayesian Optimization}, 
      author={Peter I. Frazier},
      year={2018},
      eprint={1807.02811},
      archivePrefix={arXiv},
      primaryClass={stat.ML},
      url={https://arxiv.org/abs/1807.02811}, 
}

@article{borzsonyi2013we,
  title={What we can learn about ultrashort pulses by linear optical methods},
  author={Borzsonyi, Adam and Kovacs, Attila P and Osvay, Karoly},
  journal={Applied Sciences},
  volume={3},
  number={2},
  pages={515--544},
  year={2013},
  publisher={MDPI}
}

@article{pathak2012effect,
  title={Effect of the frequency chirp on laser wakefield acceleration},
  author={Pathak, Vishwa Bandhu and Vieira, Jorge and Fonseca, RA and Silva, LO},
  journal={New Journal of Physics},
  volume={14},
  number={2},
  pages={023057},
  year={2012},
  publisher={IOP Publishing}
}

@article{toth2003tuning,
  title={Tuning of laser pulse shapes in grating-based compressors for optimal electron acceleration in plasmas},
  author={T{\'o}th, Cs and Faure, J and Van Tilborg, J and Geddes, CGR and Schroeder, CB and Esarey, E and Leemans, WP},
  journal={Optics letters},
  volume={28},
  number={19},
  pages={1823--1825},
  year={2003},
  publisher={Optical Society of America}
}

@article{grigoriadis2023efficient,
  title={Efficient plasma electron accelerator driven by linearly chirped multi-10-TW laser pulses},
  author={Grigoriadis, Anastasios and Andrianaki, Georgia and Tazes, Ioannis and Dimitriou, Vasilios and Tatarakis, Michael and Benis, EP and Papadogiannis, NA},
  journal={Scientific Reports},
  volume={13},
  number={1},
  pages={2918},
  year={2023},
  publisher={Nature Publishing Group UK London}
}

@article{shalloo2020automation,
  title={Automation and control of laser wakefield accelerators using Bayesian optimization},
  author={Shalloo, RJ and Dann, SJD and Gruse, J-N and Underwood, CID and Antoine, AF and Arran, Christopher and Backhouse, Michael and Baird, CD and Balcazar, MD and Bourgeois, Nicholas and others},
  journal={Nature communications},
  volume={11},
  number={1},
  pages={6355},
  year={2020},
  publisher={Nature Publishing Group UK London}
}

@article{paye1995space,
  title={Space--time Wigner functions and their application to the analysis of a pulse shaper},
  author={Paye, J{\'e}r{\^o}me and Migus, Arnold},
  journal={Journal of the Optical Society of America B},
  volume={12},
  number={8},
  pages={1480--1490},
  year={1995},
  publisher={Optical Society of America}
}

@article{cooley1965algorithm,
  title={An algorithm for the machine calculation of complex Fourier series},
  author={Cooley, James W and Tukey, John W},
  journal={Mathematics of computation},
  volume={19},
  number={90},
  pages={297--301},
  year={1965},
  publisher={JSTOR}
}

@article{shannon2006communication,
  title={Communication in the presence of noise},
  author={Shannon, Claude E},
  journal={Proceedings of the IRE},
  volume={37},
  number={1},
  pages={10--21},
  year={2006},
  publisher={IEEE}
}

@article{kostyukov2004phenomenological,
  title={Phenomenological theory of laser-plasma interaction in “bubble” regime},
  author={Kostyukov, I and Pukhov, A and Kiselev, S},
  journal={Physics of Plasmas},
  volume={11},
  number={11},
  pages={5256--5264},
  year={2004},
  publisher={American Institute of Physics}
}

@article{lu2007generating,
  title={Generating multi-GeV electron bunches using single stage laser wakefield acceleration<? format?> in a 3D nonlinear regime},
  author={Lu, Wei and Tzoufras, M and Joshi, C and Tsung, FS and Mori, WB and Vieira, J and Fonseca, RA and Silva, LO},
  journal={Physical Review Special Topics—Accelerators and Beams},
  volume={10},
  number={6},
  pages={061301},
  year={2007},
  publisher={APS}
}

@article{siegman1973hermite,
  title={Hermite--Gaussian functions of complex argument as optical-beam eigenfunctions},
  author={Siegman, Anthony E},
  journal={Journal of the Optical society of America},
  volume={63},
  number={9},
  pages={1093--1094},
  year={1973},
  publisher={Optical Society of America}
}

@article{allen1992orbital,
  title={Orbital angular momentum of light and the transformation of Laguerre-Gaussian laser modes},
  author={Allen, Les and Beijersbergen, Marco W and Spreeuw, RJC and Woerdman, JP},
  journal={Physical review A},
  volume={45},
  number={11},
  pages={8185},
  year={1992},
  publisher={APS}
}

@article{lifschitz2009particle,
  title={Particle-in-cell modelling of laser--plasma interaction using Fourier decomposition},
  author={Lifschitz, Agustin F and Davoine, Xavier and Lefebvre, Erik and Faure, J{\'e}r{\^o}me and Rechatin, Cl{\'e}ment and Malka, Victor},
  journal={Journal of Computational Physics},
  volume={228},
  number={5},
  pages={1803--1814},
  year={2009},
  publisher={Elsevier}
}

@article{audet2018gas,
  title={Gas cell density characterization for laser wakefield acceleration},
  author={Audet, TL and Lee, P and Maynard, G and Dufr{\'e}noy, S Dobosz and Maitrallain, A and Bougeard, M and Monot, P and Cros, B},
  journal={Nuclear Instruments and Methods in Physics Research Section A: Accelerators, Spectrometers, Detectors and Associated Equipment},
  volume={909},
  pages={383--386},
  year={2018},
  publisher={Elsevier}
}

@article{chen2012theory,
  title={Theory of ionization-induced trapping in laser-plasma accelerators},
  author={Chen, M and Esarey, E and Schroeder, CB and Geddes, CGR and Leemans, WP},
  journal={Physics of Plasmas},
  volume={19},
  number={3},
  pages={033101},
  year={2012},
  publisher={American Institute of Physics}
}

@book{steel1960principles,
  title={Principles and procedures of statistics.},
  author={Steel, Robert George Douglas and Torrie, James Hiram},
  year={1960},
  publisher={McGraw Hill, Inc}
}

@book{lawson1995solving,
  title={Solving least squares problems},
  author={Lawson, Charles L and Hanson, Richard J},
  year={1995},
  publisher={SIAM}
}

@article{kirchen2021optimal,
  title={Optimal beam loading in a laser-plasma accelerator},
  author={Kirchen, Manuel and Jalas, S{\"o}ren and Messner, Philipp and Winkler, Paul and Eichner, Timo and H{\"u}bner, Lars and H{\"u}lsenbusch, Thomas and Jeppe, Laurids and Parikh, Trupen and Schnepp, Matthias and others},
  journal={Physical review letters},
  volume={126},
  number={17},
  pages={174801},
  year={2021},
  publisher={APS}
}

@Article{couperus2017demonstration,
  author    = {Couperus, JP and Pausch, R and K{\"o}hler, A and Zarini, O and Kr{\"a}mer, JM and Garten, M and Huebl, A and Gebhardt, R and Helbig, U and Bock, S and others},
  journal   = {Nature communications},
  title     = {Demonstration of a beam loaded nanocoulomb-class laser wakefield accelerator},
  year      = {2017},
  number    = {1},
  pages     = {1--7},
  volume    = {8},
  publisher = {Nature Publishing Group},
  relevance = {relevant},
}

@article{moulanier2023modeling,
  title={Modeling of the driver transverse profile for laser wakefield electron acceleration at APOLLON research facility},
  author={Moulanier, Ioaquin and Dickson, LT and Ballage, Charles and Vasilovici, Ovidiu and Gremaud, Aubin and Dobosz Dufr{\'e}noy, Sandrine and Delerue, Nicolas and Bernardi, Lorenzo and Mahjoub, Ali and Cauchois, Antoine and others},
  journal={Physics of Plasmas},
  volume={30},
  number={5},
  year={2023},
  publisher={AIP Publishing}
}

@book{andrews1998special,
  title={Special functions of mathematics for engineers},
  author={Andrews, Larry C},
  volume={49},
  year={1998},
  publisher={Spie Press}
}

@article{Backhouse2025,
  title = {Spectral phase control for optimized ionization injection in laser wakefield acceleration},
  author = {Backhouse, M. P. and Dickson, L. T. and Moulanier, I. and Massimo, F. and Cobo, C. C. and Filippi, F. and Gustafsson, C. and Lofquist, E. and Svendsen, K. and Streeter, M. J. V. and Shalloo, R. J. and Vasilovici, O. and Ballage, C. and Dann, S. J. D. and Murphy, C. D. and Najmudin, Z. and Dobosz Dufr\'enoy, S. and Lundh, O. and Cros, B.},
  journal = {Phys. Rev. Accel. Beams},
  volume = {28},
  issue = {12},
  pages = {121301},
  numpages = {6},
  year = {2025},
  month = {Dec},
  publisher = {American Physical Society},
  doi = {10.1103/9gjq-htzn},
  url = {https://link.aps.org/doi/10.1103/9gjq-htzn}
}

@book{BirdsallLangdon,
  address = {New York},
  author = {Birdsall, Charles K. and Langdon, A. Bruce},
  isbn = {0750310251 9780750310253},
  publisher = {Taylor and Francis},
  refid = {70319677},
  title = {Plasma physics via computer simulation},
  year = 2005
}

@article{Esarey2009,
  title = {Physics of laser-driven plasma-based electron accelerators},
  author = {Esarey, E. and Schroeder, C. B. and Leemans, W. P.},
  journal = {Rev. Mod. Phys.},
  volume = {81},
  issue = {3},
  pages = {1229--1285},
  numpages = {0},
  year = {2009},
  month = {Aug},
  publisher = {American Physical Society},
  doi = {10.1103/RevModPhys.81.1229},
  url = {https://link.aps.org/doi/10.1103/RevModPhys.81.1229}
}

@article{TajimaDawson1979,
  title = {Laser Electron Accelerator},
  author = {Tajima, T. and Dawson, J. M.},
  journal = {Phys. Rev. Lett.},
  volume = {43},
  issue = {4},
  pages = {267--270},
  numpages = {0},
  year = {1979},
  month = {Jul},
  publisher = {American Physical Society},
  doi = {10.1103/PhysRevLett.43.267},
  url = {https://link.aps.org/doi/10.1103/PhysRevLett.43.267}
}

@article{jolly2019spectral,
  title={Spectral phase control of interfering chirped pulses for high-energy narrowband terahertz generation},
  author={Jolly, Spencer W and Matlis, Nicholas H and Ahr, Frederike and Leroux, Vincent and Eichner, Timo and Calendron, Anne-Laure and Ishizuki, Hideki and Taira, Takunori and K{\"a}rtner, Franz X and Maier, Andreas R},
  journal={Nature communications},
  volume={10},
  number={1},
  pages={2591},
  year={2019},
  publisher={Nature Publishing Group UK London}
}

@article{leemans2002electron,
  title={Electron-Yield Enhancement in a Laser-Wakefield Accelerator Driven<? format?> by Asymmetric Laser Pulses},
  author={Leemans, WP and Catravas, P and Esarey, E and Geddes, CGR and Toth, Cs and Trines, R and Schroeder, CB and Shadwick, BA and Van Tilborg, J and Faure, J},
  journal={Physical review letters},
  volume={89},
  number={17},
  pages={174802},
  year={2002},
  publisher={APS}
}

@article{popp2010all,
  title={All-optical steering of laser-wakefield-accelerated electron beams},
  author={Popp, Antonia and Vieira, Jorge and Osterhoff, Jens and Major, Zs and H{\"o}rlein, Rainer and Fuchs, Matthias and Weingartner, Raphael and Rowlands-Rees, TP and Marti, Michael and Fonseca, RA and others},
  journal={Physical review letters},
  volume={105},
  number={21},
  pages={215001},
  year={2010},
  publisher={APS}
}

@article{ferri2016effect,
  title={Effect of experimental laser imperfections on laser wakefield acceleration and betatron source},
  author={Ferri, J and Davoine, X and Fourmaux, S and Kieffer, JC and Corde, S{\'e}bastien and Ta Phuoc, K and Lifschitz, Agustin},
  journal={Scientific reports},
  volume={6},
  number={1},
  pages={27846},
  year={2016},
  publisher={Nature Publishing Group UK London}
}

@article{assmann2020eupraxia,
  title={EuPRAXIA conceptual design report},
  author={Assmann, RW and Weikum, MK and Akhter, T and Alesini, D and Alexandrova, AS and Anania, MP and Andreev, NE and Andriyash, I and Artioli, M and Aschikhin, A and others},
  journal={The European Physical Journal Special Topics},
  volume={229},
  number={24},
  pages={3675--4284},
  year={2020},
  publisher={Springer}
}

@article{andre2018control,
  title={Control of laser plasma accelerated electrons for light sources},
  author={Andr{\'e}, T and Andriyash, IA and Loulergue, A and Labat, M and Roussel, E and Ghaith, A and Khojoyan, M and Thaury, C and Vall{\'e}au, M and Briquez, F and others},
  journal={Nature communications},
  volume={9},
  number={1},
  pages={1334},
  year={2018},
  publisher={Nature Publishing Group UK London}
}

@article{Pariente2016,
	author = {Pariente, G. and Gallet, V. and Borot, A. and Gobert, O. and Qu{\'e}r{\'e}, F.},
	doi = {10.1038/nphoton.2016.140},
	id = {Pariente2016},
	isbn = {1749-4893},
	journal = {Nature Photonics},
	number = {8},
	pages = {547--553},
	title = {Space--time characterization of ultra-intense femtosecond laser beams},
	url = {https://doi.org/10.1038/nphoton.2016.140},
	volume = {10},
	year = {2016}}

@article{Akturk_2010,
	author = {Selcuk Akturk and Xun Gu and Pamela Bowlan and Rick Trebino},
	doi = {10.1088/2040-8978/12/9/093001},
	journal = {Journal of Optics},
	month = {aug},
	number = {9},
	pages = {093001},
	title = {Spatio-temporal couplings in ultrashort laser pulses},
	url = {https://dx.doi.org/10.1088/2040-8978/12/9/093001},
	volume = {12},
	year = {2010}}

@article{Jeandet22,
	author = {Antoine Jeandet and Spencer W. Jolly and Antonin Borot and Beno\^{i}t Bussi\`{e}re and Paul Dumont and Julien Gautier and Olivier Gobert and Jean-Philippe Goddet and Anthony Gonsalves and Arie Irman and Wim P. Leemans and Rodrigo Lopez-Martens and Gabriel Mennerat and Kei Nakamura and Marie Ouill\'{e} and Gustave Pariente and Moana Pittman and Thomas P\"{u}schel and Fabrice Sanson and Fran\c{c}ois Sylla and C\'{e}dric Thaury and Karl Zeil and Fabien Qu\'{e}r\'{e}},
	doi = {10.1364/OE.444564},
	journal = {Opt. Express},
	month = {Jan},
	number = {3},
	pages = {3262--3288},
	publisher = {Optica Publishing Group},
	title = {Survey of spatio-temporal couplings throughout high-power ultrashort lasers},
	url = {https://opg.optica.org/oe/abstract.cfm?URI=oe-30-3-3262},
	volume = {30},
	year = {2022}}

@article{tournois1997acousto,
  title={Acousto-optic programmable dispersive filter for adaptive compensation of group delay time dispersion in laser systems},
  author={Tournois, Pierre},
  journal={Optics communications},
  volume={140},
  number={4-6},
  pages={245--249},
  year={1997},
  publisher={Elsevier}
}

@article{quesnel1998theory,
  title={Theory and simulation of the interaction of ultraintense laser pulses with electrons in vacuum},
  author={Quesnel, Brice and Mora, Patrick},
  journal={Physical Review E},
  volume={58},
  number={3},
  pages={3719},
  year={1998},
  publisher={APS}
}

@article{pierce2023arbitrarily,
  title={Arbitrarily structured laser pulses},
  author={Pierce, Jacob R and Palastro, John P and Li, Fei and Malaca, Bernardo and Ramsey, Dillon and Vieira, Jorge and Weichman, Kathleen and Mori, Warren B},
  journal={Physical Review Research},
  volume={5},
  number={1},
  pages={013085},
  year={2023},
  publisher={APS}
}

@article{massimo2020numerical,
  title={Numerical modeling of laser tunneling ionization in particle-in-cell codes with a laser envelope model},
  author={Massimo, Francesco and Beck, Arnaud and D{\'e}rouillat, Julien and Zemzemi, Imen and Specka, Arnd},
  journal={Physical Review E},
  volume={102},
  number={3},
  pages={033204},
  year={2020},
  publisher={APS}
}

@article{irshad2023multi,
  title={Multi-objective and multi-fidelity Bayesian optimization of laser-plasma acceleration},
  author={Irshad, Faran and Karsch, Stefan and D{\"o}pp, Andreas},
  journal={Physical Review Research},
  volume={5},
  number={1},
  pages={013063},
  year={2023},
  publisher={APS}
}

@article{mirzaie2015demonstration,
  title={Demonstration of self-truncated ionization injection for GeV electron beams},
  author={Mirzaie, M and Li, S and Zeng, M and Hafz, NAM and Chen, M and Li, GY and Zhu, QJ and Liao, H and Sokollik, T and Liu, F and others},
  journal={Scientific reports},
  volume={5},
  number={1},
  pages={1--9},
  year={2015},
  publisher={Nature Publishing Group}
}

@Article{pak2010injection,
  author    = {Pak, Arthur and Marsh, KA and Martins, SF and Lu, W and Mori, WB and Joshi, C},
  journal   = {Physical Review Letters},
  title     = {Injection and trapping of tunnel-ionized electrons into laser-produced wakes},
  year      = {2010},
  number    = {2},
  pages     = {025003},
  volume    = {104},
  comment   = {Ionisation injection description specifically using nitrogen and helium atoms, discusses the a0 values required for self-trapping and discusses the a0 required in previous experiments to achieve this},
  publisher = {APS},
}

@Article{mcguffey2010ionization,
  author    = {McGuffey, C and Thomas, AGR and Schumaker, W and Matsuoka, T and Chvykov, V and Dollar, FJ and Kalintchenko, G and Yanovsky, V and Maksimchuk, A and Krushelnick, K and others},
  journal   = {Physical Review Letters},
  title     = {Ionization induced trapping in a laser wakefield accelerator},
  year      = {2010},
  number    = {2},
  pages     = {025004},
  volume    = {104},
  publisher = {APS},
}

@article{maier2020decoding,
  title={Decoding sources of energy variability in a laser-plasma accelerator},
  author={Maier, Andreas R and Delbos, Niels M and Eichner, Timo and H{\"u}bner, Lars and Jalas, S{\"o}ren and Jeppe, Laurids and Jolly, Spencer W and Kirchen, Manuel and Leroux, Vincent and Messner, Philipp and others},
  journal={Physical Review X},
  volume={10},
  number={3},
  pages={031039},
  year={2020},
  publisher={APS}
}

@article{Dickson2022,
  title = {Mechanisms to control laser-plasma coupling in laser wakefield electron acceleration},
  author = {Dickson, L. T. and Underwood, C. I. D. and Filippi, F. and Shalloo, R. J. and Svensson, J. Bj\"orklund and Gu\'enot, D. and Svendsen, K. and Moulanier, I. and Dufr\'enoy, S. Dobosz and Murphy, C. D. and Lopes, N. C. and Rajeev, P. P. and Najmudin, Z. and Cantono, G. and Persson, A. and Lundh, O. and Maynard, G. and Streeter, M. J. V. and Cros, B.},
  journal = {Phys. Rev. Accel. Beams},
  volume = {25},
  issue = {10},
  pages = {101301},
  numpages = {12},
  year = {2022},
  month = {Oct},
  publisher = {American Physical Society},
  doi = {10.1103/PhysRevAccelBeams.25.101301},
  url = {https://link.aps.org/doi/10.1103/PhysRevAccelBeams.25.101301}
}

@Article{gerchberg1972practical,
  author  = {Gerchberg, Ralph W},
  journal = {Optik},
  title   = {A practical algorithm for the determination of phase from image and diffraction plane pictures},
  year    = {1972},
  pages   = {237--246},
  volume  = {35},
}

@phdthesis{zemzemi2020high,
  title={High-performance computing and numerical simulation for laser wakefield acceleration with realistic laser profiles},
  author={Zemzemi, Imene},
  year={2020},
  school={Polytechnic Institute of Paris}
}

@phdthesis{moulanier2024modelisation,
  title={Mod{\'e}lisation r{\'e}aliste de l'acc{\'e}l{\'e}ration laser-plasma},
  author={Moulanier, Ioaquin},
  year={2024},
  school={Universit{\'e} Paris-Saclay}
}

@article{verluise2000amplitude,
  title={Amplitude and phase control of ultrashort pulses by use of an acousto-optic programmable dispersive filter: pulse compression and shaping},
  author={Verluise, Frederic and Laude, Vincent and Cheng, Z and Spielmann, Ch and Tournois, Pierre},
  journal={Optics letters},
  volume={25},
  number={8},
  pages={575--577},
  year={2000},
  publisher={Optical Society of America}
}

@article{kaplan2002dazzler,
author = {Kaplan, Daniel and Tournois, Pascal},
year = {2002},
month = {06},
pages = {69-75},
title = {Theory and performance of the acousto optic programmable dispersive filter used for femtosecond laser pulse shaping},
volume = {12},
journal = {Journal de Physique IV},
doi = {10.1051/jp4:20020098}
}

@article{mohseni2018resolution,
  title={Resolution of spectral focusing in coherent Raman imaging},
  author={Mohseni, Mojtaba and Polzer, Christoph and Hellerer, Thomas},
  journal={Optics express},
  volume={26},
  number={8},
  pages={10230--10241},
  year={2018},
  publisher={Optical Society of America}
}

@book{diels2006ultrashort,
  title={Ultrashort laser pulse phenomena},
  author={Diels, Jean-Claude and Rudolph, Wolfgang},
  year={2006},
  publisher={Elsevier}
}

@article{hong2002time,
  title={Time--frequency analysis of chirped femtosecond pulses using Wigner distribution function},
  author={Hong, K-H and Kim, J-H and Kang, Yong Hoon and Nam, Chang Hee},
  journal={Applied Physics B},
  volume={74},
  pages={s231--s236},
  year={2002},
  publisher={Springer}
}

@article{huijts2021identifying,
  title={Identifying observable carrier-envelope phase effects in laser wakefield acceleration with near-single-cycle pulses},
  author={Huijts, Julius and Andriyash, Igor A and Rovige, Lucas and Vernier, Aline and Faure, J{\'e}r{\^o}me},
  journal={Physics of Plasmas},
  volume={28},
  number={4},
  year={2021},
  publisher={AIP Publishing}
}

@article{vay2007noninvariance,
  title={Noninvariance of Space-and Time-Scale Ranges under a Lorentz Transformation<? format?> and the Implications for the Study of Relativistic Interactions},
  author={Vay, J-L},
  journal={Physical review letters},
  volume={98},
  number={13},
  pages={130405},
  year={2007},
  publisher={APS}
}

@article{yu2016enabling,
  title={Enabling Lorentz boosted frame particle-in-cell simulations of laser wakefield acceleration in quasi-3D geometry},
  author={Yu, Peicheng and Xu, Xinlu and Davidson, Asher and Tableman, Adam and Dalichaouch, Thamine and Li, Fei and Meyers, Michael D and An, Weiming and Tsung, Frank S and Decyk, Viktor K and others},
  journal={Journal of Computational Physics},
  volume={316},
  pages={747--759},
  year={2016},
  publisher={Elsevier}
}

@article{godfrey2014numerical,
  title={Numerical stability analysis of the pseudo-spectral analytical time-domain PIC algorithm},
  author={Godfrey, Brendan B and Vay, Jean-Luc and Haber, Irving},
  journal={Journal of Computational Physics},
  volume={258},
  pages={689--704},
  year={2014},
  publisher={Elsevier}
}

@article{vay2013domain,
  title={A domain decomposition method for pseudo-spectral electromagnetic simulations of plasmas},
  author={Vay, Jean-Luc and Haber, Irving and Godfrey, Brendan B},
  journal={Journal of Computational Physics},
  volume={243},
  pages={260--268},
  year={2013},
  publisher={Elsevier}
}

@article{jalas2021bayesian,
  title={Bayesian optimization of a laser-plasma accelerator},
  author={Jalas, S{\"o}ren and Kirchen, Manuel and Messner, Philipp and Winkler, Paul and H{\"u}bner, Lars and Dirkwinkel, Julian and Schnepp, Matthias and Lehe, Remi and Maier, Andreas R},
  journal={Physical review letters},
  volume={126},
  number={10},
  pages={104801},
  year={2021},
  publisher={APS}
}

@article{leroux2020description,
  title={Description of spatio-temporal couplings from heat-induced compressor grating deformation},
  author={Leroux, Vincent and Eichner, Timo and Maier, Andreas R},
  journal={Optics express},
  volume={28},
  number={6},
  pages={8257--8265},
  year={2020},
  publisher={Optical Society of America}
}

@article{sainte2017controlling,
  title={Controlling the velocity of ultrashort light pulses in vacuum through spatio-temporal couplings},
  author={Sainte-Marie, Antonin and Gobert, Olivier and Quere, Fabien},
  journal={Optica},
  volume={4},
  number={10},
  pages={1298--1304},
  year={2017},
  publisher={Optical Society of America}
}

@article{valenta2025bayesian,
  title={Bayesian optimization of electron energy from laser wakefield accelerators},
  author={Valenta, P and Esirkepov, T Zh and Ludwig, JD and Wilks, SC and Bulanov, SV},
  journal={Physical Review Accelerators and Beams},
  volume={28},
  number={9},
  pages={094601},
  year={2025},
  publisher={APS}
}

@article{pathak2018spectral,
  title={Spectral effects in the propagation of chirped laser pulses in uniform underdense plasma},
  author={Pathak, Naveen and Zhidkov, Alexei and Hosokai, Tomonao and Kodama, Ryosuke},
  journal={Physics of Plasmas},
  volume={25},
  number={1},
  year={2018},
  publisher={AIP Publishing}
}

@article{zeraouli2026exploring,
  title={Exploring high-intensity laser-driven secondary sources via high-order spectral pulse shaping for high-energy-density experiments},
  author={Zeraouli, G and Mariscal, DA and Grace, E and Swanson, KK and Djordjevic, BZ and Hill, MP and Jiang, S and Daskalova, R and Tiscareno, G and Hanggi, D and others},
  journal={Physics of Plasmas},
  volume={33},
  number={2},
  year={2026},
  publisher={AIP Publishing}
}

@article{loughran2023automated,
  title={Automated control and optimization of laser-driven ion acceleration},
  author={Loughran, Brendan and Streeter, Matthew JV and Ahmed, Hamad and Astbury, Sam and Balcazar, M and Borghesi, M and Bourgeois, N and Curry, CB and Dann, SJD and DiIorio, S and others},
  journal={High Power Laser Science and Engineering},
  volume={11},
  pages={e35},
  year={2023},
  publisher={Cambridge University Press}
}

@article{kalmykov2012laser,
  title={Laser plasma acceleration with a negatively chirped pulse: all-optical control over dark current in the blowout regime},
  author={Kalmykov, Serguei Y and Beck, Arnaud and Davoine, Xavier and Lefebvre, Erik and Shadwick, Bradley A},
  journal={New Journal of Physics},
  volume={14},
  number={3},
  pages={033025},
  year={2012},
  publisher={IOP Publishing}
}

@article{ferran2023bayesian,
  title={Bayesian optimization of laser-plasma accelerators assisted by reduced physical models},
  author={Ferran Pousa, A and Jalas, S and Kirchen, M and Martinez de la Ossa, A and Th{\'e}venet, M and Hudson, S and Larson, J and Huebl, A and Vay, J-L and Lehe, R},
  journal={Physical Review Accelerators and Beams},
  volume={26},
  number={8},
  pages={084601},
  year={2023},
  publisher={APS}
}

@article{irshad2024pareto,
  title={Pareto optimization and tuning of a laser wakefield accelerator},
  author={Irshad, F and Eberle, C and Foerster, FM and Grafenstein, K v and Haberstroh, F and Travac, E and Weisse, N and Karsch, Stefan and D{\"o}pp, Andreas},
  journal={Physical review letters},
  volume={133},
  number={8},
  pages={085001},
  year={2024},
  publisher={APS}
}

@article{jalas2023tuning,
  title={Tuning curves for a laser-plasma accelerator},
  author={Jalas, S{\"o}ren and Kirchen, M and Braun, C and Eichner, T and Gonzalez, JB and H{\"u}bner, Lars and H{\"u}lsenbusch, T and Messner, P and Palmer, G and Schnepp, M and others},
  journal={Physical Review Accelerators and Beams},
  volume={26},
  number={7},
  pages={071302},
  year={2023},
  publisher={APS}
}

@article{glenn2026characterization,
  title={Characterization and automated optimization of laser-driven proton beams from converging liquid sheet jet targets},
  author={Glenn, GD and Treffert, F and Ahmed, H and Astbury, S and Borghesi, M and Bourgeois, Nicola and Curry, CB and Dann, Stephen JD and DiIorio, S and Dover, NP and others},
  journal={Physical Review Research},
  volume={8},
  number={1},
  pages={013101},
  year={2026},
  publisher={APS}
}

@article{djordjevic2025bayesian,
  title={Bayesian Optimization of Laser-Wakefield Acceleration via Spectral Pulse Shaping},
  author={Djordjevi{\'c}, BZ and Benedetti, C and McNaughton, AD and Lehe, R and Tsai, H-E and Wilks, SC and Reagan, BA and Williams, GJ and van Tilborg, J and Schroeder, CB},
  journal={arXiv preprint arXiv:2512.09125},
  year={2025}
}

@article{schroeder2003frequency,
  title={Frequency chirp and pulse shape effects in self-modulated laser wakefield accelerators},
  author={Schroeder, CB and Esarey, E and Geddes, CGR and Toth, Cs and Shadwick, BA and Van Tilborg, J and Faure, J and Leemans, WP},
  journal={Physics of Plasmas},
  volume={10},
  number={5},
  pages={2039--2046},
  year={2003},
  publisher={American Institute of Physics}
}

@book{svelto2010principles,
  title={Principles of lasers},
  author={Svelto, Orazio and Hanna, David C},
  volume={1},
  year={2010},
  publisher={Springer}
}

@article{massimo2025laser,
  title = {Laser field reconstruction for the modeling of laser-plasma interaction in cylindrical geometry},
  author = {Massimo, F. and Moulanier, I. and Guerente, A. and Khomyshyn, O. and Masckala, M. and Steyn, T. L. and Schramm, U. and Irman, A. and Cros, B.},
  journal = {Phys. Rev. E},
  volume = {113},
  issue = {5},
  pages = {055211},
  numpages = {25},
  year = {2026},
  month = {May},
  publisher = {American Physical Society},
  doi = {10.1103/m1lt-wz5h},
  url = {https://link.aps.org/doi/10.1103/m1lt-wz5h}
}

@inproceedings{thevenet2025lasy,
  title={LASY: LAser manipulations made eaSY},
  author={Th{\'e}venet, Maxence and Andriyash, Igor A and Fedeli, Luca and Ferran Pousa, A and Huebl, Axel and Jalas, S{\"o}ren and Kirchen, Manuel and Lehe, Remi and Shalloo, Rob J and Sinn, Alexander and others},
  booktitle={Journal of Physics: Conference Series},
  volume={3124},
  number={1},
  pages={012014},
  year={2025},
  organization={IOP Publishing}
}

@book{born2013principles,
  title={Principles of optics: electromagnetic theory of propagation, interference and diffraction of light},
  author={Born, Max and Wolf, Emil},
  year={2013},
  publisher={Elsevier},
  comment = {Book referecing every bases of optical physics}
}

@article{dominici2007asymptotic,
  title={Asymptotic analysis of the Hermite polynomials from their differential--difference equation},
  author={Dominici, Diego},
  journal={Journal of Difference Equations and Applications},
  volume={13},
  number={12},
  pages={1115--1128},
  year={2007},
  publisher={Taylor \& Francis}
}

@inproceedings{moulanier2025coupled,
  title={Coupled dynamics of laser pulse and electron distribution from accurate modeling of laser wakefield accelerators},
  author={Dickson, LT},
  booktitle={Journal of Physics: Conference Series},
  volume={3124},
  number={1},
  pages={012013},
  year={2025},
  organization={IOP Publishing}
}

@article{deano2013strong,
  title={Strong and ratio asymptotics for Laguerre polynomials revisited},
  author={Deano, Alfredo and Huertas, Edmundo J and Marcell{\'a}n, Francisco},
  journal={Journal of Mathematical Analysis and Applications},
  volume={403},
  number={2},
  pages={477--486},
  year={2013},
  publisher={Elsevier}
}

@article{temme1990asymptotic,
  title={Asymptotic estimates for Laguerre polynomials},
  author={Temme, NM},
  journal={Zeitschrift f{\"u}r angewandte Mathematik und Physik ZAMP},
  volume={41},
  number={1},
  pages={114--126},
  year={1990},
  publisher={Springer}
}

@article{powell1984least,
  title={Least absolute deviations estimation for the censored regression model},
  author={Powell, James L},
  journal={Journal of econometrics},
  volume={25},
  number={3},
  pages={303--325},
  year={1984},
  publisher={Elsevier}
}

@misc{abramowitz1966handbook,
  title={Handbook of mathematical functions, with formulas, graphs, and mathematical tables},
  author={Abramowitz, Milton and Stegun, Irene A and Romain, Jacques E},
  year={1966},
  publisher={American Institute of Physics}
}

\end{document}